\documentclass[aps,pra,reprint,twocolumn,superscriptaddress,amsmath,longbibliography,nofootinbib]{revtex4-2}
\usepackage{amsmath}
\usepackage{amsfonts}
\usepackage{amssymb}
\usepackage{braket}
\usepackage{tcolorbox}
\usepackage{xcolor}
\usepackage{graphicx}
\usepackage{enumerate}

\begin{document}

\title{Simulated outperforms quantum reverse annealing in mean-field models}
\author{Christopher L. Baldwin} \email{baldw292@msu.edu}
\affiliation{Department of Physics and Astronomy, Michigan State University, East Lansing, Michigan 48824, USA}
\date{\today}

\begin{abstract}
Adiabatic reverse annealing (ARA) has been proposed as an improvement to conventional quantum annealing for solving optimization problems, in which one takes advantage of an initial guess at the solution to suppress problematic phase transitions.
Here we interpret the performance of ARA through its effects on the free energy landscape, and use the intuition gained to introduce a classical analogue to ARA termed ``simulated reverse annealing'' (SRA).
This makes it more difficult to claim that ARA provides a quantum advantage in solving a given problem, as not only must ARA succeed but the corresponding SRA must fail.
As a solvable example, we analyze how both protocols behave in the infinite-range (non-disordered) $p$-spin model.
Through both the thermodynamic phase diagrams and explicit dynamical behavior, we establish that the quantum algorithm has no advantage over its classical counterpart: SRA succeeds not only in every case where ARA does but even in a narrow range of parameters where ARA fails.
\end{abstract}

\maketitle

\section{Introduction} \label{sec:introduction}

Quantum annealing is a general procedure by which the ground state of a target Hamiltonian $H_0$ can be determined (say representing a complicated optimization problem or spin model)~\cite{Albash2018Adiabatic,Hauke2020Perspectives,Rajak2023Quantum}.
In the traditional setting, the degrees of freedom are $N$ Ising spins (``qubits''), $H_0$ is diagonal in the $\hat{\sigma}^z$ (``computational'') basis, and the spins evolve under the Hamiltonian
\begin{equation} \label{eq:quantum_annealing_Hamiltonian}
H(s) = s H_0 - (1 - s) \sum_{j=1}^N \hat{\sigma}_j^x.
\end{equation}
The practitioner slowly increases the parameter $s$ from 0 to 1, and the adiabatic theorem~\cite{Messiah1962,Jansen2007Bounds} guarantees that the spins will remain in their instantaneous ground state if this is done sufficiently slowly --- since the ground state at $s = 0$ is trivial and the ground state at $s = 1$ is that of $H_0$, this gives a means to prepare the desired state.
This is the quantum-mechanical analogue of the classical algorithm known as simulated annealing, with the transverse field (second term of Eq.~\eqref{eq:quantum_annealing_Hamiltonian}) playing a role similar to the temperature in a Monte Carlo simulation~\cite{Newman1999}.

The holy grail of quantum annealing research is a problem for which quantum annealing requires only polynomial time (in $N$) to find the ground state whereas all known classical algorithms require exponential time.
Although this continues to be a very active direction of research, a number of works have by now established that traditional quantum annealing is unlikely to provide such an advantage in generic hard optimization and spin-glass problems~\cite{Jorg2008Simple,Altshuler2010Anderson,Jorg2010FirstOrder,Young2010First,Hen2011Exponential,Farhi2012Performance,Bapst2013Quantum,Knysh2016ZeroTemperature,Baldwin2018Quantum}.
In short, the same local minima of the energy landscape that frustrate classical algorithms also tend to produce exponentially small gaps in the spectrum of Eq.~\eqref{eq:quantum_annealing_Hamiltonian}.
The adiabatic theorem asserts that the system will remain in its ground state if $s$ is varied slower than a rate scaling as a power of the gap, and thus exponentially small gaps translate to exponential time needed for quantum annealing to succeed.

Since the scaling of the gap is closely related to the presence of ground-state phase transitions --- the gap tends to be $O(1)$ within a phase, polynomially small at continuous transitions, and exponentially small at discontinuous transitions --- we can reformulate the above statement as follows: researchers hope to find a Hamiltonian $H_0$ for which Eq.~\eqref{eq:quantum_annealing_Hamiltonian} does not undergo any (or at worst only continuous) phase transitions as $s$ increases from 0 to 1, but the evidence strongly suggests that discontinuous transitions are in fact quite common in hard problems.

As a result, much recent work in the field has been devoted to more sophisticated variants of quantum annealing, in which new terms are added to Eq.~\eqref{eq:quantum_annealing_Hamiltonian} with the hope that they will allow one to circumvent discontinuous transitions~\cite{Seki2012Quantum,Hormozi2017Nonstoquastic,Susa2018Exponential,Susa2018Quantum,Albash2019Role,Adame2020Inhomogeneous,Tang2021Unconventional,Imoto2022Quantum,Kimura2023Convergence,Dadgar2025Anomalously}.
One noteworthy example is ``reverse annealing''~\cite{PerdomoOrtiz2011Study,Chancellor2017Modernizing}, which supposes that the practitioner has access to an excited state of $H_0$ (say by running a classical algorithm or initial round of traditional quantum annealing) and makes use of that additional knowledge.
There are in fact two well-studied variants of reverse annealing.
The first continues to use Eq.~\eqref{eq:quantum_annealing_Hamiltonian} as the Hamiltonian, but rather than increase $s$ monotonically from 0 to 1, the practitioner begins at $s = 1$ with the spins prepared in the excited state, reduces $s$ to a lower value (hence the ``reverse'' in the name), and then returns to $s = 1$, hopefully finding the spins in a lower-energy configuration than they began.
See Refs.~\cite{Marshall2019Power,Passarelli2020Reverse,Rocutto2021Quantum,Bando2022Breakdown,Mehta2025Unraveling} for examples.

\begin{figure*}[t]
\centering
\includegraphics[width=1.0\textwidth]{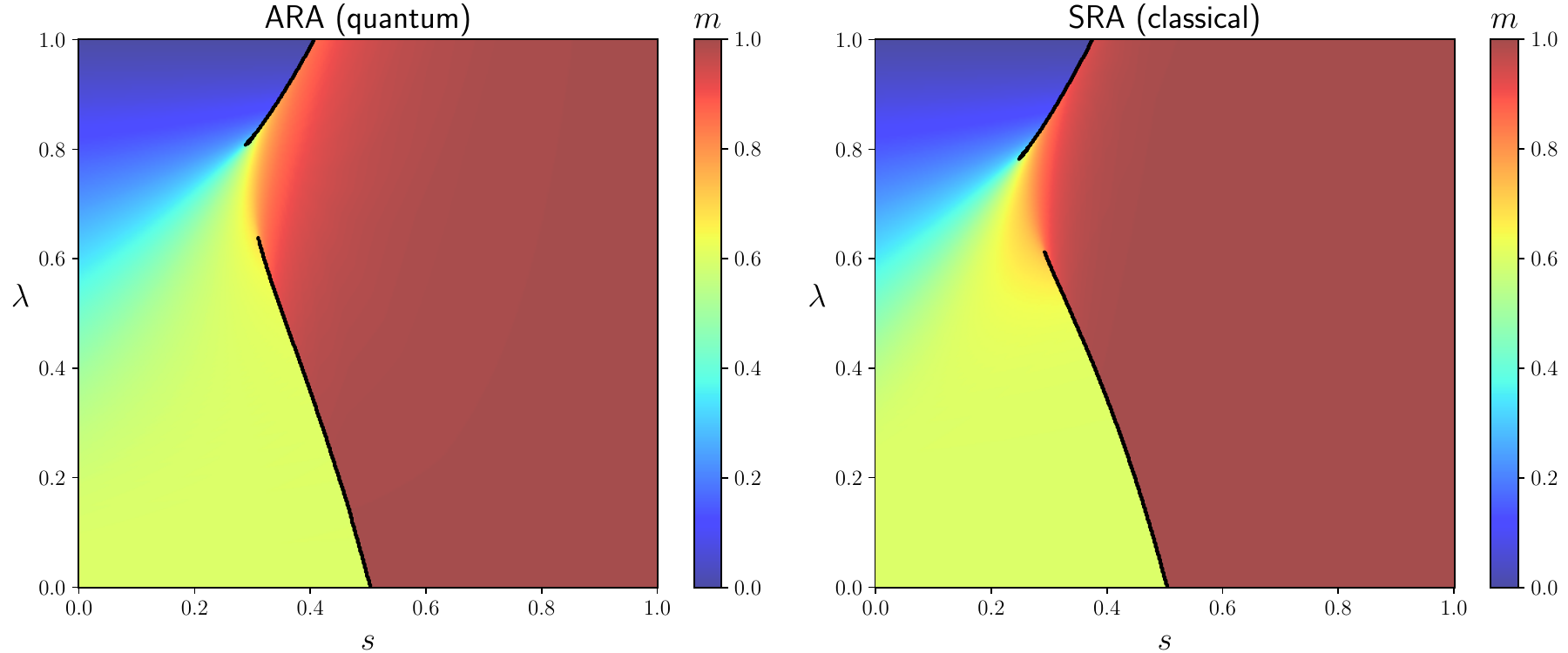}
\caption{Comparison of the equilibrium phase diagram for ARA (left) and our classical analogue SRA (right), specifically for the target Hamiltonian $H_0$ given by Eq.~\eqref{eq:Hopfield_model}, using parameter values $p = 3$, $\alpha = 0.5$, $x = 0.2$. The color indicates the equilibrium value of the magnetization $m$, with the ground state of $H_0$ corresponding to $m = 1$. Black lines indicate discontinuous phase transitions (estimated somewhat crudely as where $m$ changes by more than $0.05$ from one pixel to the next). Note that both ARA and SRA succeed for these parameter values --- there are paths from $s = 0$ to $s = 1$ that avoid phase transitions.}
\label{fig:phase_diagrams_example_success}
\end{figure*}

In this paper, however, we focus on the second variant: ``adiabatic reverse annealing'' (ARA).
Denoting the excited state by $a \equiv \{ a_j \}_{j=1}^N$ (with $a_j \in \{1, -1\}$ being the orientation of spin $j$), ARA uses the Hamiltonian
\begin{equation} \label{eq:reverse_annealing_Hamiltonian}
H(s, \lambda) = s H_0 - (1 - s) \sum_{j=1}^N \Big[ \lambda \hat{\sigma}_j^x + (1 - \lambda) a_j \hat{\sigma}_j^z \Big].
\end{equation}
The first two terms are exactly as in Eq.~\eqref{eq:quantum_annealing_Hamiltonian} --- in particular, ARA reduces to traditional quantum annealing by setting $\lambda = 1$.
The third term is a longitudinal field pointing in the direction of the excited state, such that $a$ is the ground state when $s = \lambda = 0$.
ARA begins at this point and slowly increases $s$ from 0 to 1 while simultaneously varying $\lambda$ --- the additional freedom provided by this second parameter allows for much more flexibility to find an optimal path in the $s$-$\lambda$ plane.

A number of studies of ARA have found encouraging results.
On the theoretical side, Refs.~\cite{Ohkuwa2018Reverse,Yamashiro2019Dynamics} have shown that ARA can indeed avoid phase transitions in a family of solvable mean-field models (the $p$-spin model discussed below).
The benefit even extends to more complicated problems that incorporate disorder~\cite{Arai2021Mean,Arai2022MeanField}, and counterdiabatic driving can be used to accelerate the protocol further~\cite{Passarelli2023Counterdiabatic}.
On the experimental side, Ref.~\cite{Zhang2024Cyclic} has recently run a cyclical version of ARA --- using past configurations as starting points for future iterations --- on the D-Wave quantum annealer~\cite{Johnson2011Quantum,King2022Coherent,King2023Quantum}, with noteworthy success in large-scale spin-glass problems.

\begin{figure*}[t]
\centering
\includegraphics[width=1.0\textwidth]{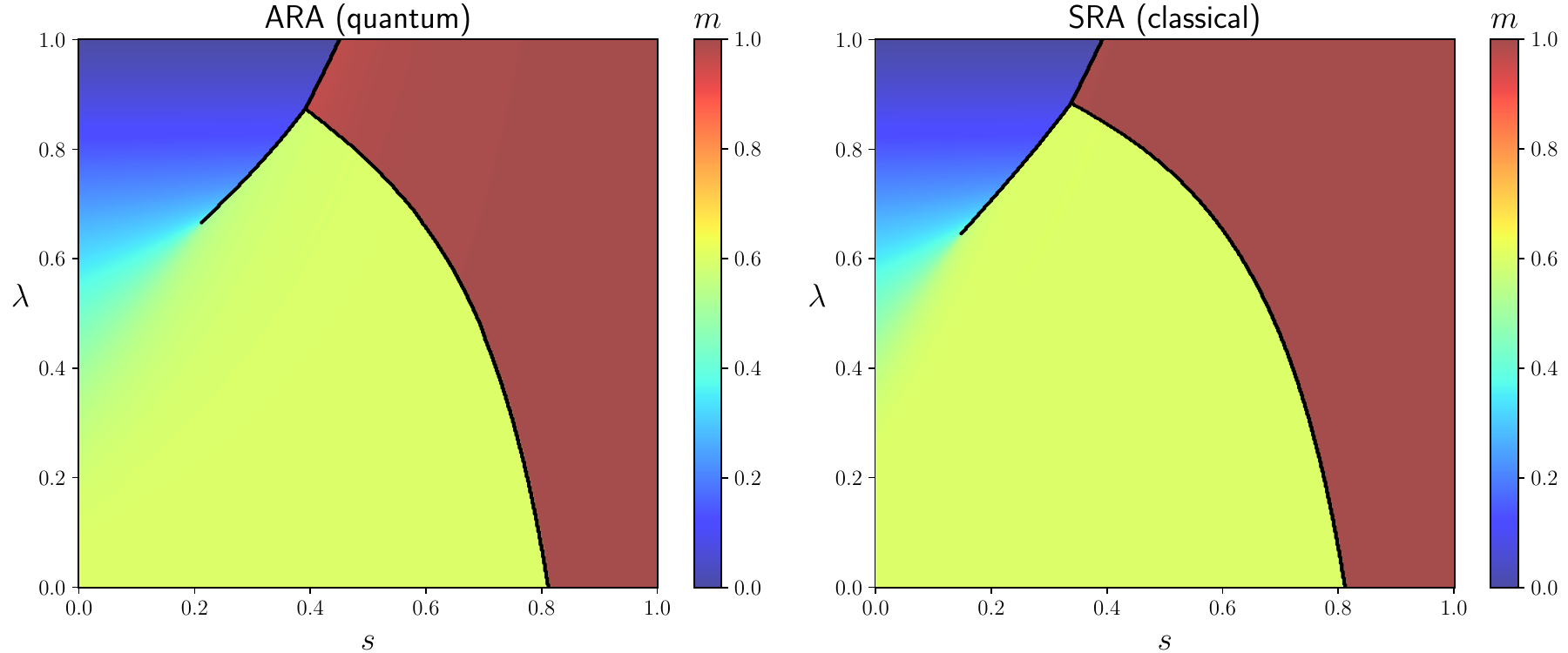}
\caption{Comparison of the phase diagram for ARA (left) and SRA (right), using parameter values $p = 5$, $\alpha = 0.9$, $x = 0.2$ (for $H_0$ given by Eq.~\eqref{eq:Hopfield_model}). The color indicates the equilibrium value of the magnetization $m$. Black lines indicate discontinuous phase transitions (again identified as where $m$ changes by more than $0.05$ from one pixel to the next). Here, in contrast to Fig.~\ref{fig:phase_diagrams_example_success}, both ARA and SRA fail --- there are no paths from $s = 0$ to $s = 1$ that do not cross a phase transition.}
\label{fig:phase_diagrams_example_failure}
\end{figure*}

To be clear, these results do not imply that ARA will always be able to circumvent phase transitions: even in solvable toy models, it only succeeds in certain parameter regimes.
Thus it is important to identify the physics underlying ARA, in particular the mechanism for suppressing discontinuous phase transitions and the conditions under which this occurs.
This is what we do in the present paper.
On the one hand, we argue that ARA could conceivably work just as well in hard spin-glass problems as in mean-field models, at least in principle (which helps to explain the experimental results of Ref.~\cite{Zhang2024Cyclic}).
Yet we also argue that quantum mechanics does not play an essential role in ARA: the success is due to fluctuations ``fattening'' the relevant wells of the free energy landscape, and any fluctuations (not merely quantum) could serve that purpose.
To make this notion more precise, and to directly demonstrate its practical implications, we introduce a classical analogue of ARA --- termed ``simulated reverse annealing'' (SRA) --- that seems to work just as well.
We show that SRA is also capable of circumventing phase transitions, in similar parameter regimes, in the $p$-spin model of Refs.~\cite{Ohkuwa2018Reverse,Yamashiro2019Dynamics} (and even a slight generalization to a two-pattern Hopfield model~\cite{Amit1985SpinGlass,Nishimori1996Quantum}).
We do so both by calculating the equilibrium phase diagram --- Figs.~\ref{fig:phase_diagrams_example_success} and~\ref{fig:phase_diagrams_example_failure} give examples showing just how similar the two are, both when they succeed and when they fail --- and by explicitly determining the dynamical behavior using mean-field techniques.

These results make it more difficult to claim that ARA provides a quantum advantage: any success case for ARA must be subjected to SRA as well, and it must be shown that the latter cannot be equally successful.
For the models that we consider here, there is a range of parameter values in which SRA succeeds and ARA fails, but no values for which the opposite holds.
In other words, SRA unambiguously outperforms ARA in mean-field models.

Although pessimistic, this should not discourage future investigations of ARA and SRA.
The two are distinct algorithms, and we see no reason why SRA should always outperform ARA in every situation (even though it does so for the toy models considered here).
Rather, there are likely certain features of the energy landscape that make one method preferable over the other, and it remains to determine what those features are.

The remainder of the paper is as follows.
We define the models under consideration in Sec.~\ref{sec:models}, then revisit ARA through the lens of the free energy landscape in Sec.~\ref{sec:free_energy_landscapes}.
We define SRA and show that it succeeds, both via thermodynamics and dynamics, in Sec.~\ref{sec:simulated_reverse_annealing}.
Lastly, we reiterate the main results and implications in Sec.~\ref{sec:conclusion}.
Details of the mean-field techniques which form the basis of our calculations are given in three appendices.

\section{Models} \label{sec:models}

A number of works~\cite{Ohkuwa2018Reverse,Yamashiro2019Dynamics,Passarelli2022Standard} have studied the performance of ARA in the (non-disordered) $p$-spin model:
\begin{equation} \label{eq:p_spin_model}
H_0 = -N \bigg( \frac{1}{N} \sum_{j=1}^N \hat{\sigma}_j^z \bigg)^p.
\end{equation}
Clearly this model is not itself a challenging optimization problem --- the ground state is immediately seen to be the all-up state in which $\hat{\sigma}_j^z = 1$ for all spins (we take $p$ odd to avoid degeneracy in the ground state).
Nonetheless, it is an important solvable toy model, as it exhibits a discontinuous phase transition under conventional annealing~\cite{Jorg2010Energy,Bapst2012On} which ARA is able to circumvent~\cite{Ohkuwa2018Reverse}.

We refer to the excited state $a$ which is input to ARA as the ``marked state'', and denote the fraction of spins that point down in the marked state by $x$.
Thus the marked state is close to the true ground state at small $x$ and far from it at large $x$.
It is convenient to define partial magnetizations
\begin{equation} \label{eq:partial_magnetization_definition}
\hat{S}_u^{\nu} \equiv \sum_j \delta_{a_j,1} \hat{\sigma}_j^{\nu}, \qquad \hat{S}_d^{\nu} \equiv \sum_j \delta_{a_j,-1} \hat{\sigma}_j^{\nu},
\end{equation}
where $\nu \in \{x, y, z\}$, so that $\hat{S}_u^{\nu}$ sums only over the spins which point up in the marked state and $\hat{S}_d^{\nu}$ sums only over the spins which point down.
Note in particular that the all-up state has $\hat{S}_u^z = N(1-x)$ and $\hat{S}_d^z = Nx$, while the marked state has $\hat{S}_u^z = N(1-x)$ and $\hat{S}_d^z = -Nx$.
The full ARA Hamiltonian (Eq.~\eqref{eq:reverse_annealing_Hamiltonian}) can be written
\begin{equation} \label{eq:p_spin_ARA_Hamiltonian}
\begin{aligned}
H(s, \lambda) = &-s N^{1-p} \big( \hat{S}_u^z + \hat{S}_d^z \big)^p \\
&- (1 - s) \Big[ \lambda \big( \hat{S}_u^x + \hat{S}_d^x \big) + (1 - \lambda) \big( \hat{S}_u^z - \hat{S}_d^z \big) \Big].
\end{aligned}
\end{equation}

Since our mean-field calculations apply to any Hamiltonian which can be written in terms of $\hat{S}_u^{\nu}$ and $\hat{S}_d^{\nu}$, we generalize the $p$-spin model to
\begin{equation} \label{eq:Hopfield_model}
\begin{aligned}
H_0 &= -N \bigg( \frac{1}{N} \sum_{j=1}^N \hat{\sigma}_j^z \bigg)^p - \alpha N \bigg( \frac{1}{N} \sum_{j=1}^N a_j \hat{\sigma}_j^z \bigg)^p \\
&= -N^{1-p} \big( \hat{S}_u^z + \hat{S}_d^z \big)^p - \alpha N^{1-p} \big( \hat{S}_u^z - \hat{S}_d^z \big)^p,
\end{aligned}
\end{equation}
with coefficient $\alpha \in (0, 1)$.
The first term favors the all-up state, while the second term favors the marked state.
Thus Eq.~\eqref{eq:Hopfield_model}, which is in effect a two-pattern Hopfield model~\cite{Amit1985SpinGlass,Nishimori1996Quantum}, is the simplest generalization to have the all-up state as the global minimum and the marked state as a local minimum.
We could in fact use any function of $\hat{S}_u^z$ and $\hat{S}_d^z$ for $H_0$, but we choose Eq.~\eqref{eq:Hopfield_model} to keep the number of parameters manageable.
The analysis that follows thus starts from the ARA Hamiltonian 
\begin{equation} \label{eq:Hopfield_ARA_Hamiltonian}
\begin{aligned}
H(s, \lambda) = &-s N^{1-p} \big( \hat{S}_u^z + \hat{S}_d^z \big)^p - s \alpha N^{1-p} \big( \hat{S}_u^z - \hat{S}_d^z \big)^p\\
&- (1 - s) \Big[ \lambda \big( \hat{S}_u^x + \hat{S}_d^x \big) + (1 - \lambda) \big( \hat{S}_u^z - \hat{S}_d^z \big) \Big].
\end{aligned}
\end{equation}

\section{Free energy landscapes} \label{sec:free_energy_landscapes}

To begin, we determine the thermodynamic phase diagram of Eq.~\eqref{eq:Hopfield_ARA_Hamiltonian} and explain its key features in terms of the free energy landscape.
ARA is concerned with the ground state, but we first compute the free energy at arbitrary temperature $T$, and take $T \rightarrow 0$ only at the end.
Since the initial calculation --- deriving a path-integral representation of the partition function solely in terms of order parameters --- is very similar to that in Ref.~\cite{Ohkuwa2018Reverse} (and many other quantum-annealing studies), we relegate details to Appendix~\ref{app:thermodynamic_path_integral}.
The only difference compared to past works is that we have two order parameters, $m_u$ and $m_d$, whose saddle-point values give the thermal expectation values of $\hat{S}_u^z/N$ and $\hat{S}_d^z/N$ respectively.

The partition function of Eq.~\eqref{eq:Hopfield_ARA_Hamiltonian} amounts to a path integral over order parameters $m_u(\tau)$ and $m_d(\tau)$, alongside their Lagrange multipliers $h_u(\tau)$ and $h_d(\tau)$, where $\tau$ is an imaginary-time index.
The integrand is of the form $\exp{[-N \beta \Phi]}$, where $\beta \equiv 1/T$ and the expression for the action $\Phi(m_u, m_d, h_u, h_d)$ is derived in Appendix~\ref{app:thermodynamic_path_integral}.
At large $N$, this path integral can be computed by saddle-point approximation, i.e., evaluating the action at points where its derivatives vanish.
Since the action is found to be invariant under shifting the imaginary-time index ($m(\tau), h(\tau) \rightarrow m(\tau + \Delta), h(\tau + \Delta)$), it is natural to look for saddle points which are independent of $\tau$ (this is often referred to as the ``static ansatz''~\cite{static_ansatz_note}).
Making this ansatz, the action is ultimately (see Appendix~\ref{app:thermodynamic_path_integral})
\begin{widetext}
\begin{equation} \label{eq:thermodynamic_static_action}
\begin{aligned}
\Phi(m_u, m_d, h_u, h_d) &= -s (m_u + m_d)^p - s \alpha (m_u - m_d)^p - (1 - s)(1 - \lambda) (m_u - m_d) + h_u m_u + h_d m_d \\
&\qquad - \frac{1-x}{\beta} \log{2 \cosh{\beta \sqrt{h_u^2 + (1-s)^2 \lambda^2}}} - \frac{x}{\beta} \log{2 \cosh{\beta \sqrt{h_d^2 + (1-s)^2 \lambda^2}}}.
\end{aligned}
\end{equation}
\end{widetext}

The normal procedure is to set all partial derivatives of $\Phi$ to zero simultaneously and solve the resulting four equations to determine the saddle points.
If there are multiple saddle points, that which gives the lowest value of the action is chosen, and within the saddle-point approximation, the free energy $f_{\textrm{eq}} \equiv -(N \beta)^{-1} \log{Z}$ is simply that lowest value.
In short,
\begin{equation} \label{eq:thermodynamic_free_energy_saddle_point_result}
f_{\textrm{eq}} = \min \Phi(m_u, m_d, h_u, h_d),
\end{equation}
where the minimization is over all saddle points.
Furthermore, as noted above, the saddle-point values of $m_u$ and $m_d$ are the thermal expectation values of $\hat{S}_u^z/N$ and $\hat{S}_d^z/N$ respectively~\cite{thermal_expectation_note}.
Thus phase transitions can be identified as points in parameter space where the saddle-point behaves non-analytically --- in particular, discontinuous transitions occur when the saddle point jumps discontinuously.

Carrying out this analysis, we obtain ARA phase diagrams very similar to those in Ref.~\cite{Ohkuwa2018Reverse}.
In particular, there are paths from $s = 0$ to $s = 1$ which avoid phase transitions when $x$ is less than a critical value (recall that $x$ is the fraction of spins pointing down in the marked state), meaning that ARA can succeed in efficiently determining the ground state of $H_0$.
The left panel of Fig.~\ref{fig:phase_diagrams_example_success} gives such an example (plotting the saddle-point value of $m \equiv m_u + m_d$ for simplicity).

Since our goal is to understand \textit{why} ARA succeeds, we calculate not only $f_{\textrm{eq}}$ but the entire free energy ``landscape''.
This is done by determining the saddle point of $\Phi$ with respect to $h_u$ and $h_d$ while holding $m_u$ and $m_d$ fixed.
The result is a function of $m_u$ and $m_d$ alone which can be interpreted as the free energy in an ensemble where $m_u$ and $m_d$ are held at fixed (potentially non-equilibrium) values.
The equilibrium values are at the global minimum of the landscape, so we can understand where and how phase transitions occur by studying how the landscape evolves as parameters are varied~\cite{landscape_subtleties_note}.

We can now take $T \rightarrow 0$ ($\beta \rightarrow \infty$).
The action becomes
\begin{widetext}
\begin{equation} \label{eq:thermodynamic_action_zero_T}
\begin{aligned}
\Phi(m_u, m_d, h_u, h_d) &= -s (m_u + m_d)^p - s \alpha (m_u - m_d)^p - (1 - s)(1 - \lambda) (m_u - m_d) + h_u m_u + h_d m_d \\
&\qquad - (1-x) \sqrt{h_u^2 + (1-s)^2 \lambda^2} - x \sqrt{h_d^2 + (1-s)^2 \lambda^2},
\end{aligned}
\end{equation}
and setting $\partial \Phi/\partial h_u = \partial \Phi/\partial h_d = 0$ gives
\begin{equation} \label{eq:thermodynamic_field_saddle_point}
m_u = \frac{(1-x) h_u}{\sqrt{h_u^2 + (1-s)^2 \lambda^2}}, \qquad m_d = \frac{x h_d}{\sqrt{h_d^2 + (1-s)^2 \lambda^2}}.
\end{equation}
Solving for $h_u$ and $h_d$, we have
\begin{equation} \label{eq:thermodynamic_field_solution}
h_u = \frac{(1-s) \lambda m_u}{\sqrt{(1-x)^2 - m_u^2}}, \qquad h_d = \frac{(1-s) \lambda m_d}{\sqrt{x^2 - m_d^2}},
\end{equation}
and inserting into the action gives the free energy landscape (for which we use the same symbol for brevity):
\begin{equation} \label{eq:free_energy_landscape}
\begin{aligned}
\Phi(m_u, m_d) &= -s (m_u + m_d)^p - s \alpha (m_u - m_d)^p - (1 - s)(1 - \lambda) (m_u - m_d) \\
&\qquad - (1 - s) \lambda \sqrt{(1-x)^2 - m_u^2} - (1 - s) \lambda \sqrt{x^2 - m_d^2}.
\end{aligned}
\end{equation}
\end{widetext}

\begin{figure}[t]
\centering
\includegraphics[width=1.0\columnwidth]{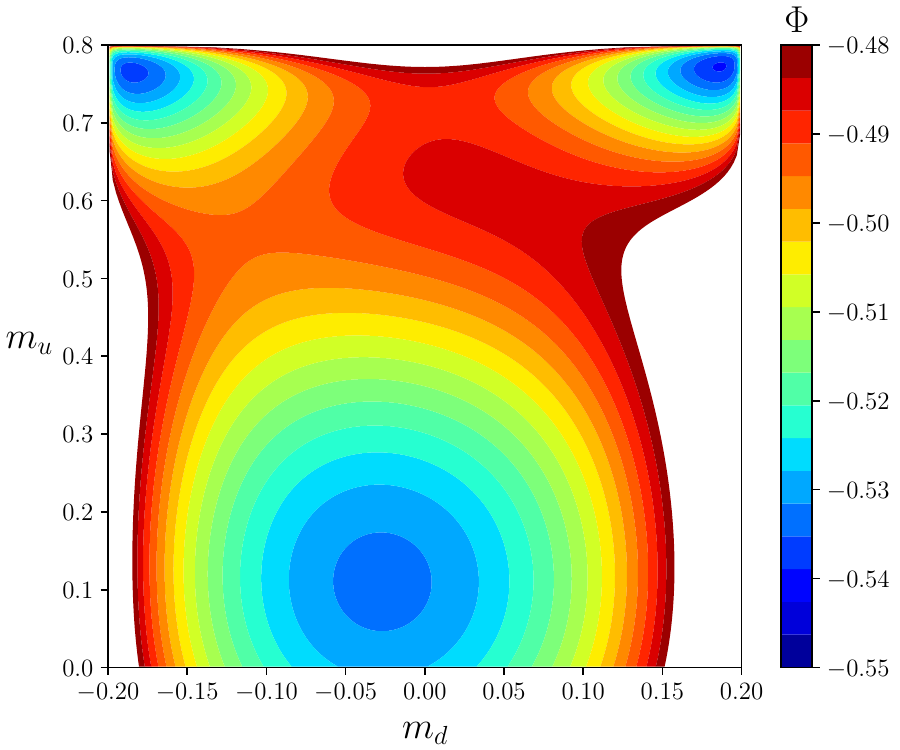}
\caption{Contour plot of the free energy landscape $\Phi(m_u, m_d)$ for $p = 5$, $\alpha = 0.9$, $x = 0.2$, at $s = 0.4$ and $\lambda = 0.88$. Only contours for $\Phi \leq -0.48$ are shown --- the white regions correspond to $\Phi(m_u, m_d) > -0.48$.}
\label{fig:quantum_landscape_example}
\end{figure}

Fig.~\ref{fig:quantum_landscape_example} gives a particularly illustrative example of the free energy landscape, in which three distinct local minima are clearly visible:
\begin{itemize}
\item One minimum is at $m_u \approx 1 - x$ and $m_d \approx x$, corresponding to states near the ground state of $H_0$ (the all-up state).
Note that the equality is inexact because of quantum fluctuations caused by the transverse field, i.e., the ground state in the presence of the field deviates from that of $H_0$~\cite{free_vs_energy_note}.
\item Another minimum is at $m_u \approx 1 - x$ and $m_d \approx -x$, corresponding to states near the marked state.
Once again, the minimum is not exactly at the marked state due to transverse-field-induced fluctuations.
\item A final minimum is at $m_u \approx 0$ and $m_d \approx 0$.
This is the paramagnetic ground state, which is realized at large transverse field.
\end{itemize}
These minima explain the discontinuous transitions seen in the phase diagram.
At small $\lambda$, the dominant terms of the Hamiltonian are $H_0$ and the longitudinal field, meaning the relevant minima of $\Phi$ are those of the all-up and marked states.
As one crosses the lower phase boundary in Fig.~\ref{fig:phase_diagrams_example_success} (say increasing $s$), the global minimum switches from the marked state to the all-up state, and thus the total magnetization $m_u + m_d$ jumps from roughly $1 - 2x$ to $1$.
Similarly, at large $\lambda$, the dominant terms are $H_0$ and the transverse field --- as one crosses the upper phase boundary, the global minimum switches from the paramagnet to the all-up state.

However, for generic parameter values, these features of the free energy landscape may not be as clear-cut.
The locations of the local minima can deviate significantly from their ideal values, certain minima may become unstable, and others may blend together.
In fact, it is this ``merging'' of minima that explains why paths at intermediate $\lambda$ can avoid phase transitions and thus why ARA can succeed, as we now show.

Using the parameter values in Fig.~\ref{fig:phase_diagrams_example_success} as an example ($p = 3$, $\alpha = 0.5$, $x = 0.2$), we study how the free energy landscape evolves along paths that begin at $(s, \lambda) = (0, 0)$ and end at $(s, \lambda) = (1, 0)$.
The initial ground state is the marked state and the final ground state is the all-up state, thus $m_d$ is the important order parameter --- it changes from $-x$ to $x$ along the path, and we compare a path where it changes discontinuously (namely following the lower edge of the phase diagram) to one where it changes smoothly (by passing through the opening seen in Fig.~\ref{fig:phase_diagrams_example_success}).
In order to better visualize changes in the landscape, we minimize over $m_u$ beforehand, i.e., we plot $\Phi'(m_d) \equiv \min_{m_u} \Phi(m_u, m_d)$.
Note that the equilibrium state is still given by the global minimum of $\Phi'$.

\begin{figure}[t]
\centering
\includegraphics[width=1.0\columnwidth]{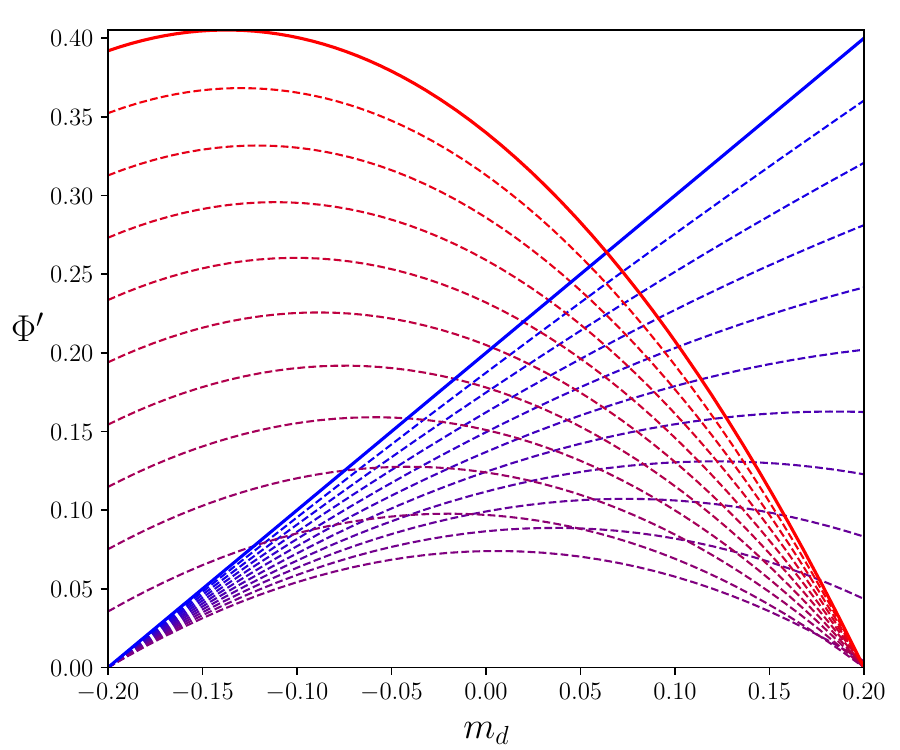}
\caption{Evolution of the free energy landscape $\Phi'(m_d)$ along the straight line from $(s, \lambda) = (0, 0)$ to $(1, 0)$ (same parameters as in Fig.~\ref{fig:phase_diagrams_example_success}) --- blue to red corresponds to increasing $s$ from 0 to 1. The $y$-axis of each curve is shifted so that the minimum always has a value of zero.}
\label{fig:quantum_first_order_path}
\end{figure}

First consider the path that increases $s$ from 0 to 1 while keeping $\lambda = 0$ throughout, i.e., without ever turning on the transverse field.
The evolution of the landscape is shown in Fig.~\ref{fig:quantum_first_order_path}.
It gives a classic example of a discontinuous phase transition --- there are two well-defined local minima at $m_d = -x$ and $x$ respectively, and which one is lower switches at a critical value of $s$.

\begin{figure}[t]
\centering
\includegraphics[width=1.0\columnwidth]{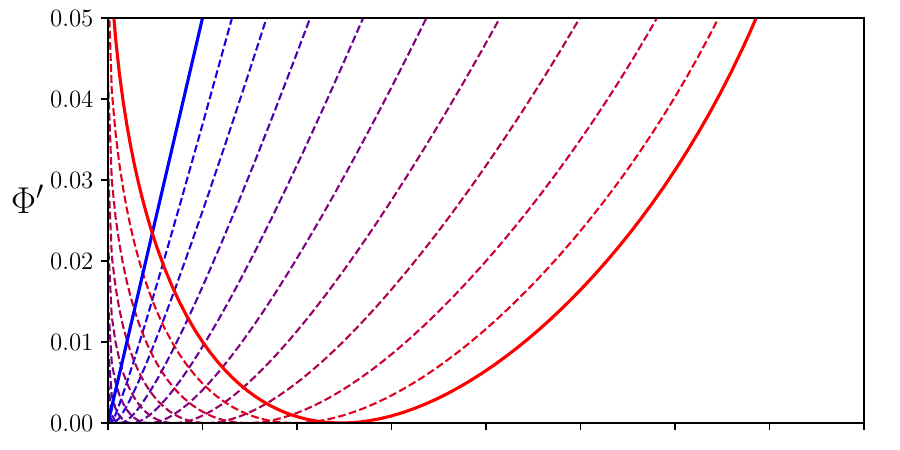}
\includegraphics[width=1.0\columnwidth]{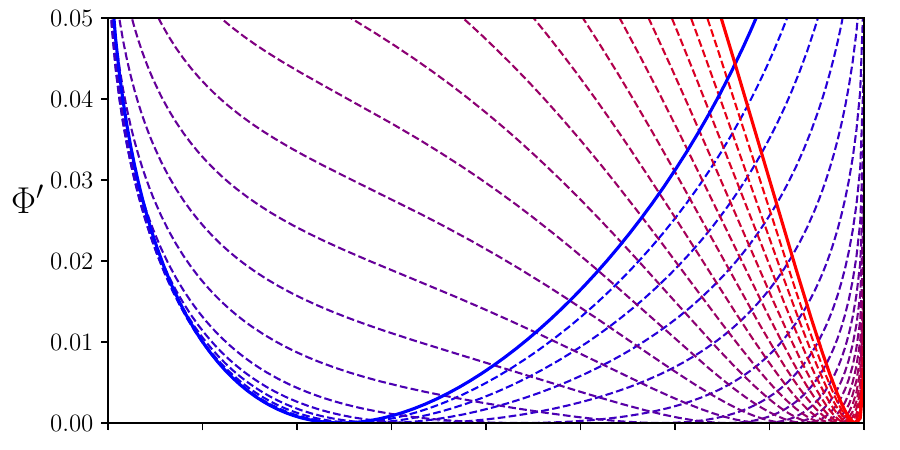}
\includegraphics[width=1.0\columnwidth]{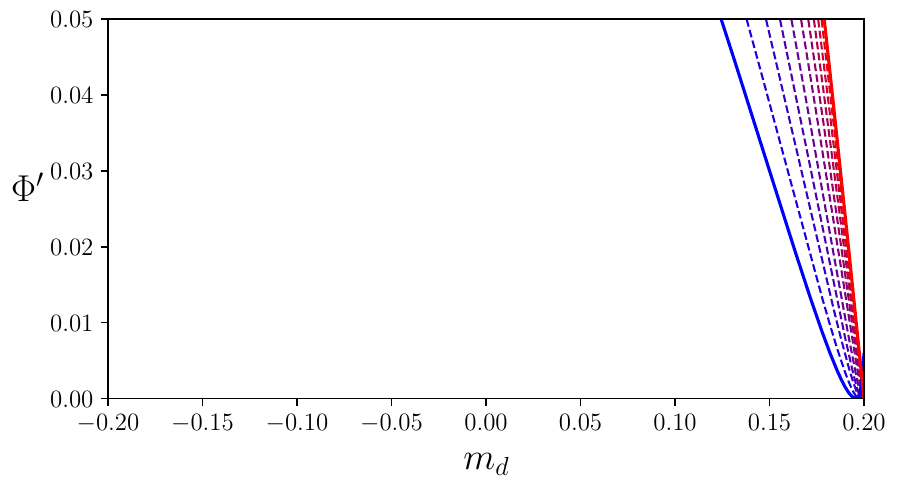}
\caption{Evolution of the free energy landscape $\Phi'(m_d)$ along a three-stage path that avoids phase transitions (same parameters as in Fig.~\ref{fig:phase_diagrams_example_success}). Top panel shows the first stage from $(s, \lambda) = (0, 0)$ to $(0.2, 0.7)$. Middle panel shows the second stage from $(0.2, 0.7)$ to $(0.6, 0.7)$. Bottom panel shows the third stage from $(0.6, 0.7)$ to $(1, 0)$. In each, blue to red corresponds to moving forward along the path (increasing $s$), and the $y$-axis of each curve is shifted so that the minimum always has a value of zero.}
\label{fig:quantum_avoiding_path}
\end{figure}

Contrast this with a path consisting of three line segments (refer to Fig.~\ref{fig:phase_diagrams_example_success} and note that this path circumvents the phase boundaries): first from $(0, 0)$ to $(0.2, 0.7)$, then from $(0.2, 0.7)$ to $(0.6, 0.7)$, then from $(0.6, 0.7)$ to $(1, 0)$.
The evolution is shown in Fig.~\ref{fig:quantum_avoiding_path}.
During the first stage, in which $\lambda$ increases substantially and the transverse field becomes strong, the minimum at $m_d = -x$ becomes much broader due to quantum fluctuations.
During the second stage, this minimum smoothly shifts towards larger $m_d$ as $s$ increases.
There is never a jump in the location of the minimum since there is never a secondary local minimum to begin with --- quantum fluctuations have blended the two together.
Lastly, during the third stage, the transverse field reduces to zero and the minimum narrows as it continues towards $m_d = x$.
Taken together, the minimum has shifted continuously from $m_d = -x$ to $x$ during the protocol.

This example demonstrates that fluctuations, by causing the local minima around the marked and all-up states to coalesce, are the key mechanism underlying ARA.
Yet it can be delicate to choose the strength of the fluctuations (governed by $\lambda$) correctly, which explains why ARA only succeeds if the marked state is sufficiently close to the all-up state.
If $\lambda$ is increased too far during the first stage, then the fluctuations drive the system into the paramagnetic state with $m_u \approx 0$, at which point increasing $s$ behaves no differently than for traditional quantum annealing.
The marked and all-up minima have to coalesce without reducing $m_u$ substantially, and this is possible only if the energy barrier separating the two is sufficiently small in length and/or height.
Fig.~\ref{fig:phase_diagrams_example_failure} gives an example in which there simply is no successful intermediate range of $\lambda$ (it is interesting to note that the system instead undergoes multiple discontinuous transitions at intermediate $\lambda$).

Lastly, note that while our analysis is limited to the $p$-spin model, nothing in the mechanism that we have identified relies on the permutation symmetry that renders the $p$-spin model solvable.
It is conceivable that ARA could succeed in genuinely hard optimization problems as well, as long as the local minima surrounding the marked state and true ground state are sufficiently close that fluctuations can merge them together without driving the system into faraway states.
Of course, one likely cannot predict beforehand what counts as ``sufficiently close'' or what paths in the $s$-$\lambda$ plane would be successful in a given problem --- one would have to search by trial-and-error, but this is always the case when applying heuristic methods to hard problems.

\section{Simulated reverse annealing} \label{sec:simulated_reverse_annealing}

Our analysis suggests that there is nothing particularly special about the \textit{quantum} fluctuations induced by a transverse field in ARA --- other sources of fluctuations may work just as well.
This motivates us to consider a ``simulated'' reverse annealing (SRA) which uses the thermal fluctuations induced by finite temperature in place of a transverse field.
Algorithmically, the SRA protocol is nothing more than running a (classical) Monte Carlo simulation in which the temperature $T$ is varied in time as the transverse field is for ARA.
For simplicity, here we consider SRA which uses a simple Metropolis update scheme --- flipping a spin with probability $\min \{ e^{-\Delta E/T}, 1 \}$, where $\Delta E$ is the change in energy upon flipping the spin --- using the Hamiltonian
\begin{equation} \label{eq:simulated_reverse_annealing_Hamiltonian}
H(s, \lambda) = s H_0 - (1 - s)(1 - \lambda) \sum_{j=1}^N a_j \sigma_j,
\end{equation}
and temperature
\begin{equation} \label{eq:simulated_reverse_annealing_temperature}
T(s, \lambda) = (1 - s) \lambda.
\end{equation}
Note that Eq.~\eqref{eq:simulated_reverse_annealing_Hamiltonian} is a \textit{classical} Hamiltonian involving classical bits $\sigma_j \in \{ +1, -1 \}$.
The user initializes the simulation in the marked state with $s = \lambda = 0$, then slowly increases $s$ from 0 to 1 while simultaneously varying $\lambda$ as the simulation proceeds.
If successful, the spins will be found in the ground state of $H_0$ at the end.

\subsection{Thermodynamics} \label{subsec:thermodynamics}

Just as the performance of ARA is tied to the ground-state phase diagram of Eq.~\eqref{eq:reverse_annealing_Hamiltonian}, so is the performance of SRA tied to the finite-temperature phase diagram of Eq.~\eqref{eq:simulated_reverse_annealing_Hamiltonian} (parametrizing the temperature as in Eq.~\eqref{eq:simulated_reverse_annealing_temperature}).
In particular, discontinuous phase transitions are problematic.
According to the standard Landau theory of phase transitions, discontinuous transitions occur when the global minimum of the free energy landscape switches from one local minimum to another.
Both local minima are stable on either side of the transition, and there is a free energy barrier separating the two even at the transition point.
Thus even though the equilibrium state changes suddenly at the transition, the system can only follow that state after the time needed to thermally activate over the barrier.
In mean-field models such as the $p$-spin model, the free energy barrier is $O(N)$ and thus the activation timescale (assuming a simple Arrhenius form) is exponential in $N$.
The same is true in \textit{disordered} infinite-range models, which often are genuinely difficult problems, and the timescale can be extremely large even in short-range models.
Thus to summarize, SRA is efficient only if there are paths in the $s$-$\lambda$ plane that avoid discontinuous transitions, exactly as for ARA.

Since we calculated the action of the transverse-field model at finite temperature (Eq.~\eqref{eq:thermodynamic_static_action}), we do not need to start over to study the thermodynamics of SRA --- simply neglect the term $(1-s) \lambda$, coming from the transverse field, inside the square roots in the lower line of Eq.~\eqref{eq:thermodynamic_static_action}.
Instead set $T = \beta^{-1} = (1 - s) \lambda$, so that the action is
\begin{widetext}
\begin{equation} \label{eq:thermodynamic_action_zero_field}
\begin{aligned}
\Phi(m_u, m_d, h_u, h_d) &= -s (m_u + m_d)^p - s \alpha (m_u - m_d)^p - (1 - s)(1 - \lambda) (m_u - m_d) + h_u m_u + h_d m_d \\
&\qquad - (1 - s) \lambda (1 - x) \log{2 \cosh{\frac{h_u}{(1 - s) \lambda}}} - (1 - s) \lambda x \log{2 \cosh{\frac{h_d}{(1 - s) \lambda}}}.
\end{aligned}
\end{equation}
We can again determine the free energy landscape.
Setting $\partial \Phi/\partial h_u = \partial \Phi/\partial h_d = 0$ gives
\begin{equation} \label{eq:SRA_field_saddle_point}
m_u = (1 - x) \tanh{\frac{h_u}{(1 - s) \lambda}}, \qquad m_d = x \tanh{\frac{h_d}{(1 - s) \lambda}},
\end{equation}
for which the solution is
\begin{equation} \label{eq:SRA_field_solution}
h_u = \frac{(1 - s) \lambda}{2} \log{\frac{1 - x + m_u}{1 - x - m_u}}, \qquad h_d = \frac{(1 - s) \lambda}{2} \log{\frac{x + m_d}{x - m_d}}.
\end{equation}
Inserting back into the action, the free energy landscape is thus
\begin{equation} \label{eq:SRA_free_energy_landscape}
\begin{aligned}
\Phi(m_u, m_d) &= -s (m_u + m_d)^p - s \alpha (m_u - m_d)^p - (1 - s)(1 - \lambda) (m_u - m_d) \\
&\qquad \qquad \qquad + (1 - s) \lambda \bigg[ \frac{1 - x + m_u}{2} \log{(1 - x + m_u)} + \frac{1 - x - m_u}{2} \log{(1 - x - m_u)} \\
&\qquad \qquad \qquad \qquad \qquad \qquad+ \frac{x + m_d}{2} \log{(x + m_d)} + \frac{x - m_d}{2} \log{(x - m_d)} \bigg],
\end{aligned}
\end{equation}
\end{widetext}
where we have neglected a constant term.
Once again, we determine the phase diagram by minimizing $\Phi(m_u, m_d)$: the location of the minimum gives the thermal expectation values of $\hat{S}_u^z/N$ and $\hat{S}_d^z/N$, so we identify discontinuous phase transitions by where there is a jump in the global minimum.

\begin{figure*}[t]
\centering
\includegraphics[width=1.0\textwidth]{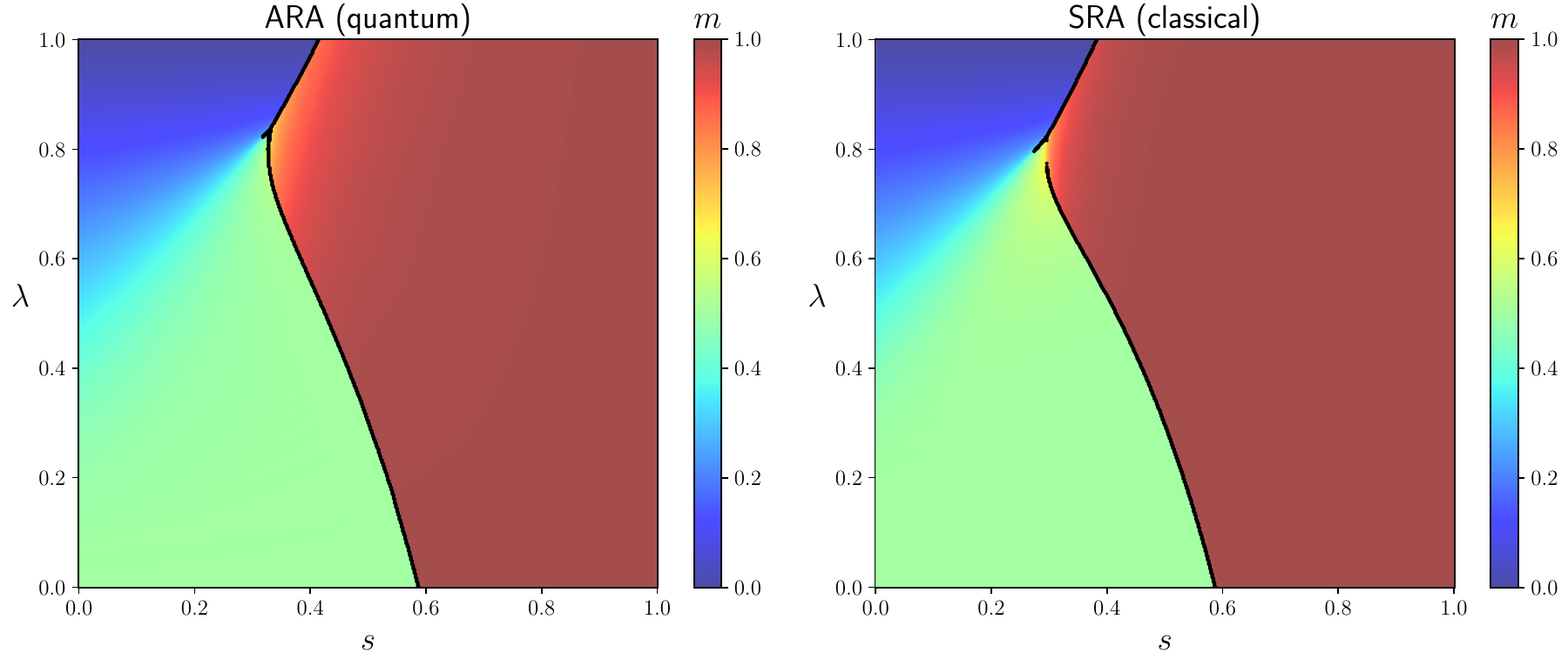}
\caption{Comparison of the phase diagram for ARA (left) and SRA (right), using parameter values $p = 3$, $\alpha = 0.6$, $x = 0.25$. The color indicates the equilibrium value of $m \equiv m_u + m_d$. Black lines indicate discontinuous phase transitions. Here, in contrast to both Figs.~\ref{fig:phase_diagrams_example_success} and~\ref{fig:phase_diagrams_example_failure}, ARA fails but SRA allows for a narrow window of successful paths.}
\label{fig:phase_diagrams_competition}
\end{figure*}

The right panel of Fig.~\ref{fig:phase_diagrams_example_success} shows the SRA phase diagram for the same parameters as in the left panel for ARA.
The two are strikingly similar (although not identical) --- there are paths from $s = 0$ to $s = 1$ at intermediate $\lambda$ which avoid phase transitions, and even the numerical values of the allowed $\lambda$ are quite close ($\lambda$ approximately between $0.6$ and $0.8$).
We observe the same trend in which the range of allowed $\lambda$ decreases as $x$ increases and disappears altogether beyond a critical value.
Similarly, the right panel of Fig.~\ref{fig:phase_diagrams_example_failure} compares the SRA phase diagram to that of ARA in a case where they both fail, and once again, the features of the two are remarkably alike.
The fact that SRA behaves so analogously to ARA demonstrates that the quantum origin of fluctuations in the latter is indeed irrelevant (and by the same token, there is nothing special about the thermal origin of fluctuations in the former).

Since ARA and SRA are both capable of locating the desired ground state in principle, the remaining question is whether there are parameter values in which one succeeds but not the other.
We have scanned across the various parameters of the model ($p$, $\alpha$, $x$) and not found any instance in which ARA succeeds while SRA fails.
Interestingly, there is a small sliver of instances in which SRA (barely) succeeds while ARA fails --- Fig.~\ref{fig:phase_diagrams_competition} gives an example.
Thus we can say that SRA unambiguously outperforms ARA in the mean-field models considered here.
Of course, since these models are not themselves interesting optimization problems, the true question --- how ARA and SRA compare in hard problems --- remains open.

\subsection{Dynamics} \label{subsec:dynamics}

Lastly, it is worthwhile to explicitly study the dynamical behavior of both protocols, as confirmation that their phase diagrams do accurately predict their performance.
We do so via path-integral calculations analogous to that of the thermodynamics.
One subtlety is that our calculations only give the expectation values $m_u(t) \equiv \langle \hat{S}_u^z(t) \rangle/N$ and $m_d(t) \equiv \langle \hat{S}_d^z(t) \rangle/N$ in the thermodynamic limit (where $\langle \, \cdot \, \rangle$ denotes the average in the time-evolved quantum state for ARA and the average over runs of the Monte Carlo simulation for SRA).
In particular, they do not give the probability of measuring the spins to be in the ground state of $H_0$.
Nonetheless, the average energy density is determined by $m_u(t)$ and $m_d(t)$ alone:
\begin{equation} \label{eq:time_evolved_energy_density}
\frac{\langle H_0(t) \rangle}{N} = -\big( m_u(t) + m_d(t) \big)^p - \alpha \big( m_u(t) - m_d(t) \big)^p.
\end{equation}
Thus to be more precise, we are studying how ARA and SRA perform in \textit{approximate} optimization --- we determine whether the measured spin configuration at the end of each protocol will (as a function of runtime $\tau$) have an energy within a finite-percentage error of the ground-state value.
In particular, if $m_u(\tau) \approx 1 - x$ and $m_d(\tau) \approx x$, then the error is small.

The details of the path-integral calculations are given in Appendices~\ref{app:ARA_path_integral} and~\ref{app:SRA_path_integral}.
While the intermediate expressions are somewhat cumbersome, the final results are simple and intuitive.
For both ARA and SRA, due to the all-to-all nature of the interactions in $H_0$, the interacting many-body dynamics can be reduced to each spin evolving in the ``mean field'' produced by the other spins but with no explicit interactions.
In the $p$-spin model considered in previous works, where all spins are equivalent due to the permutation symmetry, this means simulating a single spin with the mean field determined self-consistently from its magnetization.
In our case, where we only have permutation symmetry within two subsets of spins (those that point up in the marked state and those that point down), we must simulate two spins --- call them spin $u$ and spin $d$ --- evolving under separate self-consistent fields.
For both ARA and SRA, those mean fields are
\begin{equation} \label{eq:dynamics_self_consistent_up_field}
\begin{aligned}
h_u(t) &= s(t) p \big( m_u(t) + m_d(t) \big)^{p-1} \\
&\qquad \qquad + s(t) \alpha p \big( m_u(t) - m_d(t) \big)^{p-1} \\
&\qquad \qquad \qquad \qquad + \big( 1 - s(t) \big) \big( 1 - \lambda(t) \big),
\end{aligned}
\end{equation}
\begin{equation} \label{eq:dynamics_self_consistent_down_field}
\begin{aligned}
h_d(t) &= s(t) p \big( m_u(t) + m_d(t) \big)^{p-1} \\
&\qquad \qquad - s(t) \alpha p \big( m_u(t) - m_d(t) \big)^{p-1} \\
&\qquad \qquad \qquad \qquad - \big( 1 - s(t) \big) \big( 1 - \lambda(t) \big).
\end{aligned}
\end{equation}
While $m_u(t)$ and $m_d(t)$ still give the expectation values of $\hat{S}_u^z(t)/N$ and $\hat{S}_d^z(t)/N$ in the original problem, self-consistency requires that $m_u(t) = (1-x) \langle \hat{\sigma}_u^z(t) \rangle$ and $m_d(t) = x \langle \hat{\sigma}_d^z(t) \rangle$ for the single spins $u$ and $d$ as well.

The procedure for determining $m_u(t)$ and $m_d(t)$ is thus as follows (see Appendices~\ref{app:ARA_path_integral} and~\ref{app:SRA_path_integral} for precise statements of the algorithms):
\begin{itemize}
\item For ARA, consider a spin $u$ with wavefunction $| \psi_u(t) \rangle$ and spin $d$ with wavefunction $| \psi_d(t) \rangle$.
Since we study protocols beginning at $s = \lambda = 0$, initially $| \psi_u(0) \rangle = | \uparrow \; \rangle$ and $| \psi_d(0) \rangle = | \downarrow \; \rangle$.
Once $| \psi_u(t) \rangle$ and $| \psi_d(t) \rangle$ have been computed up to time $t$, first take the expectation values of $\hat{\sigma}^z$ (times $1-x$ and $x$ respectively) to determine $m_u(t)$ and $m_d(t)$.
Then evolve $| \psi_u(t) \rangle$ for a short time $\Delta t$ in transverse field $(1 - s(t)) \lambda(t)$ and longitudinal field $h_u(t)$ (given by Eq.~\eqref{eq:dynamics_self_consistent_up_field}), and similarly evolve $| \psi_d(t) \rangle$ in transverse field $(1 - s(t)) \lambda(t)$ and longitudinal field $h_d(t)$ (Eq.~\eqref{eq:dynamics_self_consistent_down_field}).
This determines $| \psi_u(t + \Delta t) \rangle$ and $| \psi_d(t + \Delta t) \rangle$.
Repeat.
\item For SRA, consider a spin $u$ with (classical) probability distribution $\rho_u(\sigma; t)$ and spin $d$ with distribution $\rho_d(\sigma; t)$.
Initially set $\rho_u(\sigma; 0) = \delta_{\sigma, 1}$ and $\rho_d(\sigma; 0) = \delta_{\sigma, -1}$.
Once $\rho_u(\sigma; t)$ and $\rho_d(\sigma; t)$ have been computed up to time $t$, again take the expectation values of $\sigma$ (times $1-x$ and $x$) to determine $m_u(t)$ and $m_d(t)$.
Eqs.~\eqref{eq:dynamics_self_consistent_up_field} and~\eqref{eq:dynamics_self_consistent_down_field} then give $h_u(t)$ and $h_d(t)$, and these fix the probabilities that the spins will flip in a short time $\Delta t$ according to the Metropolis update rule.
The spin-flip probabilities in turn determine $\rho_u(\sigma; t + \Delta t)$ and $\rho_d(\sigma; t + \Delta t)$.
Repeat.
\end{itemize}

\begin{figure}[t]
\centering
\includegraphics[width=1.0\columnwidth]{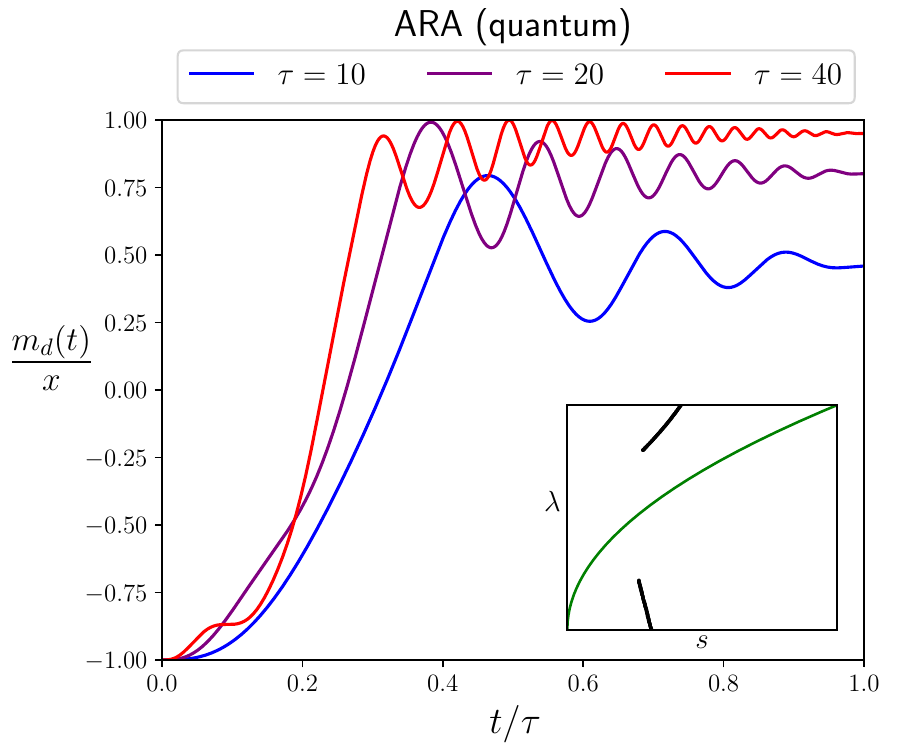}
\includegraphics[width=1.0\columnwidth]{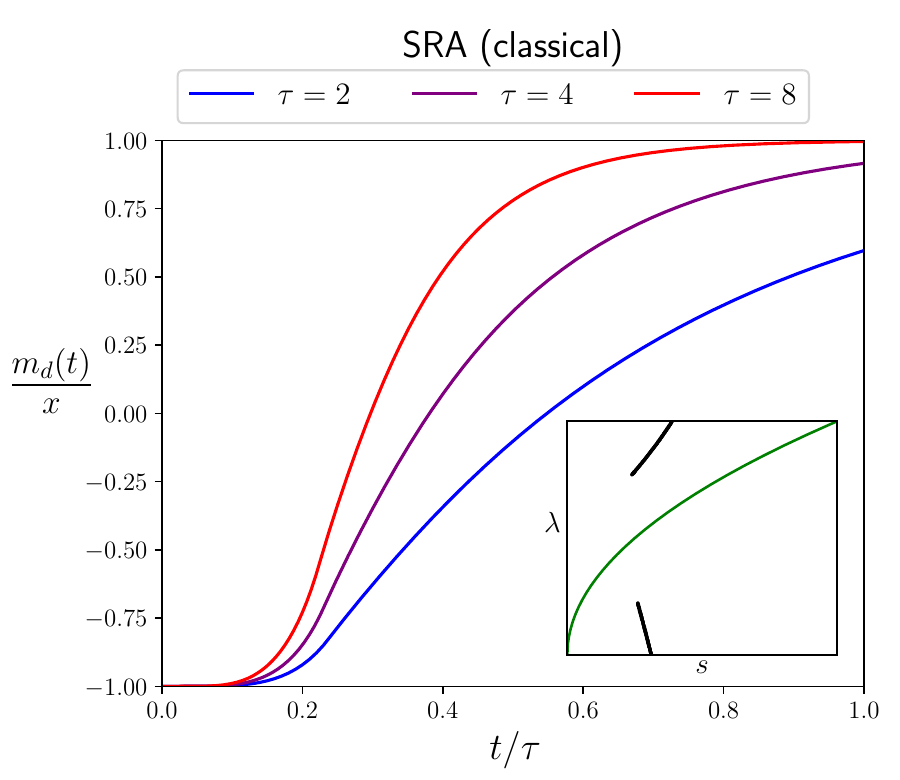}
\caption{Trajectory of $m_d(t)$ during representative ARA (top) and SRA (bottom) protocols that avoid phase transitions. Parameter values for both are $p = 3$, $\alpha = 0.1$, $x = 0.1$, following the path $s(t) = t/\tau$, $\lambda(t) = \sqrt{t/\tau}$ (different $\tau$ given by different colors). Each inset indicates the path as a green line in the corresponding phase diagram, with discontinuous transitions indicated by black lines.}
\label{fig:dynamics_examples_success}
\end{figure}

Fig.~\ref{fig:dynamics_examples_success} gives a representative example of $m_d(t)$ during a protocol that avoids phase transitions, for both ARA (top) and SRA (bottom).
Each protocol begins in the marked state, hence $m_d(0) = -x$, and one hopes to find that $m_d(\tau) \approx x$ by the end.
Both protocols indeed accomplish this, with the final value of $m_d$ coming closer to $x$ as the runtime $\tau$ increases ($m_u$ is not shown but remains close to $1-x$ throughout each protocol).
Furthermore, since the path-integral calculations implicitly take $N \rightarrow \infty$ before $\tau \rightarrow \infty$, all choices of $\tau$ are automatically $O(1)$ with respect to $N$.
Fig.~\ref{fig:dynamics_examples_success} thus demonstrates that ARA and SRA are both capable of locating the desired ground state in $O(1)$ time when phase transitions can be avoided, within an error that decreases to zero as $\tau$ increases (while still remaining $O(1)$).

Beyond this, the most noticeable difference between the two is that the classical protocol can reach the desired state in much less time than the quantum protocol --- compare the values of $\tau$ used in Fig.~\ref{fig:dynamics_examples_success}.
This makes sense: temperature-induced fluctuations directly drive the system towards a local minimum of the free energy, whereas Hamiltonian dynamics cause the spins to (at least within a semiclassical picture) precess about their local fields without relaxing.
Note that we also observe this in the oscillations that exist for ARA but not SRA.
Since the key effect at the heart of the protocols is removing barriers in the free energy landscape, it thus stands to reason that SRA should be able to reach the global minimum more efficiently.

\begin{figure}[t]
\centering
\includegraphics[width=1.0\columnwidth]{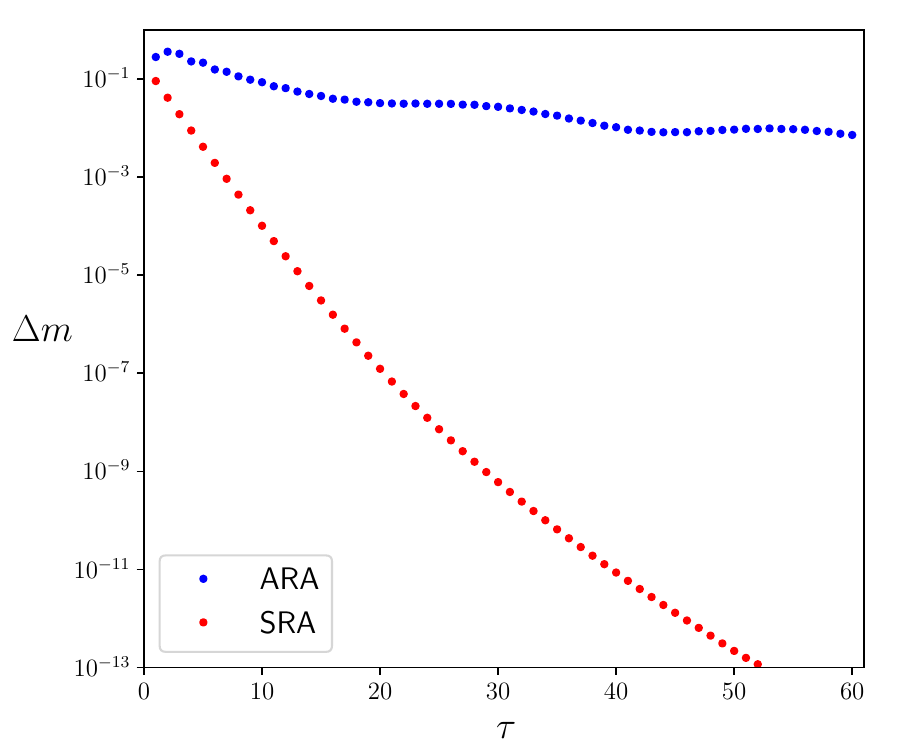}
\caption{Deviation of the magnetization from its desired value (i.e., $1$) at the end of the protocol, $\Delta m \equiv 1 - m_u(\tau) - m_d(\tau)$, as a function of runtime $\tau$. Parameter values are the same as in Fig.~\ref{fig:dynamics_examples_success} ($p = 3$, $\alpha = 0.1$, $x = 0.1$, following the path $s(t) = t/\tau$, $\lambda(t) = \sqrt{t/\tau}$).}
\label{fig:final_state_comparison}
\end{figure}

Indeed, the difference in timescale between ARA and SRA becomes even more dramatic as $\tau$ increases.
Fig.~\ref{fig:final_state_comparison} plots the magnetization at the end of each protocol, expressed as the difference $\Delta m \equiv 1 - m_u(\tau) - m_d(\tau)$ from the desired value of $1$, as a function of $\tau$.
While it is difficult to extract a precise scaling with $\tau$ in both cases --- for ARA because of slight oscillations in $\Delta m$, and for SRA because $\Delta m$ quickly reaches machine precision --- it is clear that the error decays much more rapidly for SRA.
Thus in situations where both ARA and SRA avoid phase transitions, the latter is the preferred method (although of course the bigger advantage is that one does not need a quantum annealer to run SRA).

\begin{figure}[t]
\centering
\includegraphics[width=1.0\columnwidth]{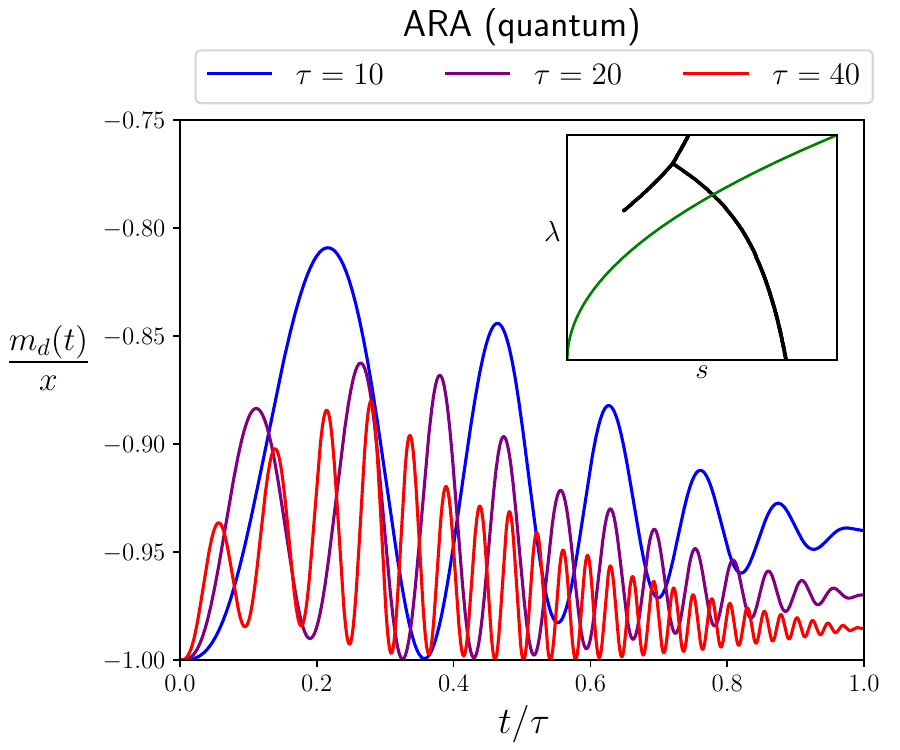}
\includegraphics[width=1.0\columnwidth]{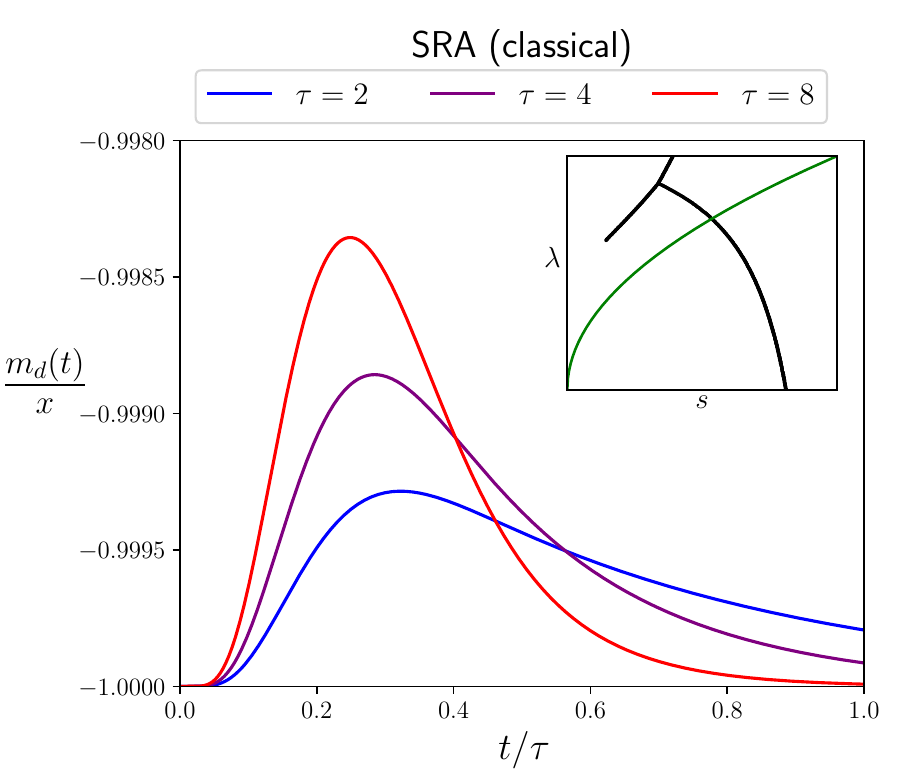}
\caption{Trajectory of $m_d(t)$ during representative ARA (top) and SRA (bottom) protocols that cross phase transitions. Parameter values for both are $p = 5$, $\alpha = 0.9$, $x = 0.2$, following the path $s(t) = t/\tau$, $\lambda(t) = \sqrt{t/\tau}$ (different $\tau$ given by different colors). Each inset indicates the path as a green line in the corresponding phase diagram, with discontinuous transitions indicated by black lines.}
\label{fig:dynamics_examples_failure}
\end{figure}

Lastly, for completeness, we also consider an example in which each protocol does cross a phase transition.
As seen in Fig.~\ref{fig:dynamics_examples_failure}, here $m_d(t)$ remains close to $-x$ throughout the protocol (note in particular the $y$-axis scale for SRA).
Moreover, the final value $m_d(\tau)$ moves even closer to $-x$ as $\tau$ increases --- clearly the protocol is failing to reach the desired state.
This makes perfect sense in terms of the free energy landscape, since the minimum around the initial marked state remains locally stable throughout the protocol~\cite{spinodal_note}.

\section{Conclusion} \label{sec:conclusion}

We have introduced a classical analogue to adiabatic reverse annealing (ARA), termed simulated reverse annealing (SRA), which is nothing more than a classical Monte Carlo simulation that varies the temperature in the same manner as one does the transverse field for ARA.
Using a slight generalization of the (non-disordered) $p$-spin model as an example, we compare the parameter regimes in which each protocol succeeds in avoiding phase transitions, and find that the two behave extremely similarly (see Figs.~\ref{fig:phase_diagrams_example_success} and~\ref{fig:phase_diagrams_example_failure}).
We explain this by analyzing how ARA modifies the free energy landscape of the model --- the transverse field (assuming its strength is neither too small nor too large) introduces fluctuations that merge the local and global minima together, allowing one to smoothly interpolate from the marked state to the true ground state by increasing $s$.
Importantly, this mechanism does not rely on the symmetries that make the $p$-spin model solvable, suggesting that it could in principle work for hard problems as well.
Yet it also shows that the origin of the fluctuations is irrelevant.
Other types of fluctuations, such as those built into a Monte Carlo simulation, may be just as effective, and the striking similarity between the ARA and SRA phase diagrams demonstrates this.

We have also directly studied the dynamics of the protocols, and confirmed that both do indeed reach the desired ground state efficiently when following paths that avoid phase transitions.
Note that we use a slightly different figure of merit than many previous works --- by using the final magnetization as a proxy, we are studying the time $\tau$ needed for the measured configuration to have energy within an $O(1)$ percentage of the ground state with high probability.
Our results confirm that this time is $O(1)$ for both protocols (when avoiding transitions).
That said, the precise value of $\tau$ is significantly smaller for SRA than for ARA.

Of course, since ARA requires an actual quantum annealer whereas SRA can be performed on any personal computer, the latter is preferable wherever it succeeds.
The important question is thus whether there is any problem for which ARA succeeds while SRA does not.
In the toy models considered here, we have unfortunately not found any such cases, and even identified a sliver for which SRA succeeds while ARA fails.
We leave the study of more sophisticated problems for future work, although this would necessarily be quite difficult since these are (by intention) problems whose low-energy properties are hard to compute.

Regardless of theoretical justification, SRA is a readily implementable classical algorithm, so it can be applied to large-scale optimization problems and its performance investigated numerically.
This is a worthwhile direction for future research.
It would be particularly interesting to run an SRA version of the recent ARA experiment that was done using the D-Wave annealer~\cite{Zhang2024Cyclic}, to see whether it performs similarly well in large spin-glass models.

Another direction for future work is to study the effects of \textit{simultaneous} quantum and thermal fluctuations in reverse annealing, as opposed to one or the other as we have done here.
For one thing, this is relevant for practical implementations of ARA, which will not be at strictly zero temperature.
Ref.~\cite{Passarelli2022Standard} has studied ARA in an open-system setting with precisely this in mind, and the concurrent Ref.~\cite{Le2025Adiabatic} takes this further by determining (among other things) the ARA phase diagram of the $p$-spin model at finite temperature.
Yet it would also be interesting to investigate whether any benefit comes from treating $T$ as a third tunable parameter in addition to $s$ and $\lambda$ --- perhaps there are regimes in which both ARA and SRA fail individually, yet there are more complicated paths in the $s$-$\lambda$-$T$ plane which do circumvent phase transitions.

Lastly, we mention that one could conceivably carry out an analogous investigation of ARA and SRA in the canonical mean-field spin-glass models (such as the \textit{disordered} $p$-spin model)~\cite{Cugliandolo1999RealTime,Castellani2005Spin,Mezard2009} --- adding a term to the Hamiltonian that couples to a specific spin configuration, as reverse annealing does with the marked state, is reminiscent of the Franz-Parisi potential from the spin-glass literature~\cite{Franz1995Recipes}.
It would be worthwhile to explore whether ideas from that field can offer insight on the performance of reverse annealing.

\section{Acknowledgements} \label{sec:acknowledgements}

This work has been supported by the U.S. National Science Foundation under award No.~2508604, as well as start-up funds from Michigan State University.

\appendix

\section{Thermodynamic path integral} \label{app:thermodynamic_path_integral}

Here we give details of the path-integral calculation for computing the thermodynamic free energy and its ``landscape''.
Subsequent appendices give the analogous derivations for the dynamics of ARA and SRA.
By presenting them side by side, we aim to highlight the common structure underlying all three.

The thermodynamic partition function is, noting that $\hat{S}_u^z + \hat{S}_d^z = \sum_j \hat{\sigma}_j^z$ and $\hat{S}_u^z - \hat{S}_d^z = \sum_j a_j \hat{\sigma}_j^z$,
\begin{widetext}
\begin{equation} \label{eq:thermodynamic_partition_function_v1}
Z = \textrm{Tr} \exp{\left[ N \beta s \bigg( \frac{1}{N} \sum_j \hat{\sigma}_j^z \bigg)^p + N \beta s \alpha \bigg( \frac{1}{N} \sum_j a_j \hat{\sigma}_j^z \bigg)^p + \beta (1 - s)(1 - \lambda) \sum_j a_j \hat{\sigma}_j^z + \beta (1 - s) \lambda \sum_j \hat{\sigma}_j^x \right]},
\end{equation}
where $\beta \equiv 1/T$ is the inverse temperature.
Using a standard Suzuki-Trotter decomposition~\cite{Trotter1959On,Suzuki1976Relationship,Sachdev2011}, we replace the operators $\hat{\sigma}_j^z$ by classical variables $\sigma_j(\tau) \in \{1, -1\}$ at the expense of introducing an extra ``imaginary-time'' label $\tau$:
\begin{equation} \label{eq:thermodynamic_partition_function_v2}
Z = \sum_{\{ \sigma_j(\tau) \}} \exp{\left[ N \int_0^{\beta} \textrm{d}\tau \left( s \bigg( \frac{1}{N} \sum_j \sigma_j(\tau) \bigg)^p + s \alpha \bigg( \frac{1}{N} \sum_j a_j \sigma_j(\tau) \bigg)^p + \frac{(1 - s)(1 - \lambda)}{N} \sum_j a_j \sigma_j(\tau) \right) + \sum_j H_{\textrm{TF},j} \right]},
\end{equation}
where $H_{\textrm{TF},j}$, coming from the transverse field, is an interaction between the same spin at different imaginary times but not between different spins (the precise form is not necessary here but can be found in standard texts~\cite{Sachdev2011}).
Next introduce $\delta$-functions which fix $\sum_j \delta_{a_j,1} \sigma_j(\tau)$ to equal $Nm_u(\tau)$ and $\sum_j \delta_{a_j,-1} \sigma_j(\tau)$ to equal $Nm_d(\tau)$:
\begin{equation} \label{eq:thermodynamic_partition_function_v3}
\begin{aligned}
Z &= \int \prod_{\tau} \textrm{d}m_u(\tau) \textrm{d}m_d(\tau) \exp{\Bigg[ N \int_0^{\beta} \textrm{d}\tau \Big( s \big( m_u(\tau) + m_d(\tau) \big)^p + s \alpha \big( m_u(\tau) - m_d(\tau) \big)^p + (1 - s)(1 - \lambda) \big( m_u(\tau) - m_d(\tau) \big) \Big) \Bigg]} \\
&\qquad \qquad \cdot \sum_{\{ \sigma_j(\tau) \}} \prod_{\tau} \delta \bigg( N m_u(\tau) - \sum_j \delta_{a_j,1} \sigma_j(\tau) \bigg) \delta \bigg( N m_d(\tau) - \sum_j \delta_{a_j,-1} \sigma_j(\tau) \bigg) \exp{\bigg[ \sum_j H_{\textrm{TF},j} \bigg]},
\end{aligned}
\end{equation}
and express the $\delta$-functions as integrals of complex exponentials ($\delta(x) \propto \int_{-i \infty}^{i \infty} \textrm{d}h \, e^{hx}$):
\begin{equation} \label{eq:thermodynamic_partition_function_v4}
\begin{aligned}
Z &\propto \int \prod_{\tau} \textrm{d}m_u(\tau) \textrm{d}m_d(\tau) \int_{-i \infty}^{i \infty} \prod_{\tau} \textrm{d}h_u(\tau) \textrm{d}h_d(\tau) \exp{\Bigg[ N \int_0^{\beta} \textrm{d}\tau \Big( s \big( m_u(\tau) + m_d(\tau) \big)^p + s \alpha \big( m_u(\tau) - m_d(\tau) \big)^p} \\
&\qquad \qquad \qquad \qquad \qquad \qquad \qquad \qquad \qquad \qquad + (1 - s)(1 - \lambda) \big( m_u(\tau) - m_d(\tau) \big) - h_u(\tau) m_u(\tau) - h_d(\tau) m_d(\tau) \Big) \Bigg] \\
&\qquad \qquad \cdot \sum_{\{ \sigma_j(\tau) \}} \exp{\Bigg[ \sum_j \bigg( \int_0^{\beta} \textrm{d}\tau \Big( \delta_{a_j,1} h_u(\tau) \sigma_j(\tau) + \delta_{a_j,-1} h_d(\tau) \sigma_j(\tau) \Big) + H_{\textrm{TF},j} \bigg) \Bigg]},
\end{aligned}
\end{equation}
where we have omitted unimportant prefactors.
The sum over spin configurations now factors over $j$, and in fact, there are only two distinct factors depending on whether $a_j = 1$ or $-1$:
\begin{equation} \label{eq:thermodynamic_single_spin_partition_function}
Z_u \equiv \sum_{\{ \sigma(\tau) \}} \exp{\Bigg[ \int_0^{\beta} \textrm{d}\tau \, h_u(\tau) \sigma(\tau) + H_{\textrm{TF}} \Bigg]}, \qquad Z_d \equiv \sum_{\{ \sigma(\tau) \}} \exp{\Bigg[ \int_0^{\beta} \textrm{d}\tau \, h_d(\tau) \sigma(\tau) + H_{\textrm{TF}} \Bigg]}.
\end{equation}
The bottom line of Eq.~\eqref{eq:thermodynamic_partition_function_v4} is simply $Z_u^{N(1-x)} Z_d^{Nx}$.
Thus the partition function amounts to the path integral of $\exp{[-N \beta \Phi(m_u, m_d, h_u, h_d)]}$, where the (Euclidean) action $\Phi$ is
\begin{equation} \label{eq:thermodynamic_action}
\begin{aligned}
\Phi(m_u, m_d, h_u, h_d) &= -\frac{1}{\beta} \int_0^{\beta} \textrm{d}\tau \Big( s \big( m_u(\tau) + m_d(\tau) \big)^p + s \alpha \big( m_u(\tau) - m_d(\tau) \big)^p + (1 - s)(1 - \lambda) \big( m_u(\tau) - m_d(\tau) \big) \Big) \\
&\qquad + \frac{1}{\beta} \int_0^{\beta} \textrm{d}\tau \Big( h_u(\tau) m_u(\tau) + h_d(\tau) m_d(\tau) \Big) - \frac{1-x}{\beta} \log{Z_u(h_u)} - \frac{x}{\beta} \log{Z_d(h_d)}.
\end{aligned}
\end{equation}
At large $N$, we can calculate this path integral by saddle-point approximation, i.e., evaluating the action at points where its derivatives vanish.
Since the action is invariant under shifting the imaginary-time index ($m(\tau), h(\tau) \rightarrow m(\tau + \Delta), h(\tau + \Delta)$), we make the ``static ansatz''~\cite{static_ansatz_note} and look for saddle points which are independent of $\tau$.
Then $Z_u$ (similarly $Z_d$) can be computed exactly, as it becomes the Suzuki-Trotter decomposition for a single spin in longitudinal field $h_u$ (similarly $h_d$) and the original transverse field $(1-s) \lambda$:
\begin{equation} \label{eq:thermodynamic_single_spin_partition_function_evaluated}
\begin{aligned}
Z_{u (d)} &= \textrm{Tr} \exp{\Big[ \beta h_{u (d)} \hat{\sigma}^z + \beta (1-s) \lambda \hat{\sigma}^x \Big]} \\
&= 2 \cosh{\beta \sqrt{h_{u (d)}^2 + (1-s)^2 \lambda^2}}.
\end{aligned}
\end{equation}
The action within the static ansatz is thus
\begin{equation} \label{eq:thermodynamic_static_action_repeat}
\begin{aligned}
\Phi(m_u, m_d, h_u, h_d) &= -s (m_u + m_d)^p - s \alpha (m_u - m_d)^p - (1 - s)(1 - \lambda) (m_u - m_d) \\
&\qquad + h_u m_u + h_d m_d - \frac{1-x}{\beta} \log{2 \cosh{\beta \sqrt{h_u^2 + (1-s)^2 \lambda^2}}} - \frac{x}{\beta} \log{2 \cosh{\beta \sqrt{h_d^2 + (1-s)^2 \lambda^2}}}.
\end{aligned}
\end{equation}
\end{widetext}

\section{ARA dynamical path integral} \label{app:ARA_path_integral}

We analyze the real-time dynamics of the ARA protocol analogously to the thermodynamics, simply starting from the Keldysh generating functional instead of the partition function.
We still use the Hamiltonian given by Eq.~\eqref{eq:Hopfield_ARA_Hamiltonian}, but now $s$ and $\lambda$ are functions of time $t \in [0, \tau]$ ($\tau$ is the total runtime of the protocol).
To simplify notation, we write $H(t)$ for the time-dependent Hamiltonian rather than the more precise $H(s(t), \lambda(t))$.
The Keldysh generating functional is then defined as
\begin{equation} \label{eq:Keldysh_generating_functional_definition}
\mathcal{Z} = \textrm{Tr} \Big( \mathcal{T} e^{-i \int_0^{\tau} \textrm{d}t H(t)} \Big) \rho \Big( \mathcal{T} e^{-i \int_0^{\tau} \textrm{d}t H(t)} \Big)^{\dag},
\end{equation}
where $\rho$ is the density matrix of the initial state and $\mathcal{T}$ denotes time-ordering.
We assume the initial state is the marked state $a$, although the calculation works for any product state $\rho = \prod_j \rho_j$.

Since $\rho$ is normalized ($\textrm{Tr} \rho = 1$) and the dynamics is unitary, in fact $\mathcal{Z} = 1$.
Nonetheless, it is very useful to construct a path-integral representation of $\mathcal{Z}$ and evaluate it by saddle-point approximation, since the location of the saddle point gives the expectation values $\langle \hat{S}_u^z(t) \rangle/N$ and $\langle \hat{S}_d^z(t) \rangle/N$, analogously to the thermodynamic path integral.
More precisely, the expectation value of $\hat{S}_u^z$ (similarly $\hat{S}_d^z$) at time $t$ is
\begin{equation} \label{eq:Keldysh_generating_functional_expectation}
\big< \hat{S}_u^z(t) \big> = \textrm{Tr} \hat{S}_u^z \Big( \mathcal{T} e^{-i \int_0^t \textrm{d}t' H(t')} \Big) \rho \Big( \mathcal{T} e^{-i \int_0^t \textrm{d}t' H(t')} \Big)^{\dag}.
\end{equation}
During the calculation of $\mathcal{Z}$, we introduce an integration variable $m_u(t)$ which replaces $\hat{S}_u^z/N$ at time $t$ --- following the same steps in Eq.~\eqref{eq:Keldysh_generating_functional_expectation} thus yields the same path integral with an additional factor of $m_u(t)$ in the integrand.
According to the saddle-point approximation, this evaluates to the saddle-point value of $m_u(t)$.
For this reason, our aim in the following is to determine the saddle point for $m_u(t)$ (and similarly $m_d(t)$).

Written out explicitly, the Keldysh generating functional is
\begin{widetext}
\begin{equation} \label{eq:Keldysh_generating_functional_v1}
\begin{aligned}
\mathcal{Z} &= \textrm{Tr} \Bigg( \mathcal{T} \exp{\Bigg[ i \int_0^{\tau} \textrm{d}t \Bigg( N s(t) \bigg( \frac{1}{N} \sum_j \hat{\sigma}_j^z \bigg)^p + N s(t) \alpha \bigg( \frac{1}{N} \sum_j a_j \hat{\sigma}_j^z \bigg)^p} \\
&\qquad \qquad \qquad \qquad \qquad \qquad \qquad \qquad \qquad \qquad + \big( 1 - s(t) \big) \big( 1 - \lambda(t) \big) \sum_j a_j \hat{\sigma}_j^z + \big( 1 - s(t) \big) \lambda(t) \sum_j \hat{\sigma}_j^x \Bigg) \Bigg] \Bigg) \\
&\qquad \cdot \rho \Bigg( \mathcal{T} \exp{\Bigg[ i \int_0^{\tau} \textrm{d}t \Bigg( N s(t) \bigg( \frac{1}{N} \sum_j \hat{\sigma}_j^z \bigg)^p + N s(t) \alpha \bigg( \frac{1}{N} \sum_j a_j \hat{\sigma}_j^z \bigg)^p} \\
&\qquad \qquad \qquad \qquad \qquad \qquad \qquad \qquad \qquad \qquad + \big( 1 - s(t) \big) \big( 1 - \lambda(t) \big) \sum_j a_j \hat{\sigma}_j^z + \big( 1 - s(t) \big) \lambda(t) \sum_j \hat{\sigma}_j^x \Bigg) \Bigg] \Bigg)^{\dag}.
\end{aligned}
\end{equation}
We use the same Suzuki-Trotter decomposition, only now for each of the two time-ordered exponentials separately.
Thus the trace becomes a sum over classical variables $\sigma_j^{\omega}(t)$, where $\omega \in \{ +, - \}$ indicates the exponential ($+$ for the top line of Eq.~\eqref{eq:Keldysh_generating_functional_v1} and $-$ for the bottom line):
\begin{equation} \label{eq:Keldysh_generating_functional_v2}
\begin{aligned}
\mathcal{Z} &= \sum_{\{ \sigma_j^{\omega}(t) \}} \exp{\Bigg[ iN \int_0^{\tau} \textrm{d}t \sum_{\omega} \omega \Bigg( s(t) \bigg( \frac{1}{N} \sum_j \sigma_j^{\omega}(t) \bigg)^p + s(t) \alpha \bigg( \frac{1}{N} \sum_j a_j \sigma_j^{\omega}(t) \bigg)^p + \frac{\big( 1 - s(t) \big) \big( 1 - \lambda(t) \big)}{N} \sum_j a_j \sigma_j^{\omega}(t) \Bigg) \Bigg]} \\
&\qquad \qquad \cdot \exp{\bigg[ \sum_j H_{\textrm{TF},j} \bigg]} \prod_j \big< \sigma_j^+(0) \big| \rho_j \big| \sigma_j^-(0) \big>.
\end{aligned}
\end{equation}
Once again, $H_{\textrm{TF},j}$ couples the same spin at different times but does not couple different spins --- its precise form is unimportant since we will ultimately absorb it back into a single-spin generating functional (analogous to Eq.~\eqref{eq:thermodynamic_single_spin_partition_function_evaluated}).
We also have the boundary conditions $\sigma_j^+(\tau) = \sigma_j^-(\tau)$, but these are unimportant for the same reason.
Next introduce $\delta$-functions which fix $\sum_j \delta_{a_j,1} \sigma_j^{\omega}(t)$ to equal $Nm_u^{\omega}(t)$ and $\sum_j \delta_{a_j,-1} \sigma_j^{\omega}(t)$ to equal $Nm_d^{\omega}(t)$:
\begin{equation} \label{eq:Keldysh_generating_functional_v3}
\begin{aligned}
\mathcal{Z} &= \int \prod_{t,\omega} \textrm{d}m_u^{\omega}(t) \textrm{d}m_d^{\omega}(t) \exp{\Bigg[ iN \int_0^{\tau} \textrm{d}t \sum_{\omega} \omega \Big( s(t) \big( m_u^{\omega}(t) + m_d^{\omega}(t) \big)^p + s(t) \alpha \big( m_u^{\omega}(t) - m_d^{\omega}(t) \big)^p} \\
&\qquad \qquad \qquad \qquad \qquad \qquad \qquad \qquad \qquad \qquad \qquad \qquad \qquad  + \big( 1 - s(t) \big) \big( 1 - \lambda(t) \big) \big( m_u^{\omega}(t) - m_d^{\omega}(t) \big) \Big) \Bigg] \\
&\quad \cdot \sum_{\{ \sigma_j^{\omega}(t) \}} \prod_{t,\omega} \delta \bigg( N m_u^{\omega}(t) - \sum_j \delta_{a_j,1} \sigma_j^{\omega}(t) \bigg) \delta \bigg( N m_d^{\omega}(t) - \sum_j \delta_{a_j,-1} \sigma_j^{\omega}(t) \bigg) \exp{\bigg[ \sum_j H_{\textrm{TF},j} \bigg]} \prod_j \big< \sigma_j^+(0) \big| \rho_j \big| \sigma_j^-(0) \big>,
\end{aligned}
\end{equation}
and express the $\delta$-functions as integrals of complex exponentials:
\begin{equation} \label{eq:Keldysh_generating_functional_v4}
\begin{aligned}
\mathcal{Z} &\propto \int \prod_{t,\omega} \textrm{d}m_u^{\omega}(t) \textrm{d}m_d^{\omega}(t) \int_{-\infty}^{\infty} \prod_{t,\omega} \textrm{d}h_u^{\omega}(t) \textrm{d}h_d^{\omega}(t) \exp{\Bigg[ iN \int_0^{\tau} \textrm{d}t \sum_{\omega} \omega \Big( s(t) \big( m_u^{\omega}(t) + m_d^{\omega}(t) \big)^p + s(t) \alpha \big( m_u^{\omega}(t) - m_d^{\omega}(t) \big)^p} \\
&\qquad \qquad \qquad \qquad \qquad \qquad \qquad \qquad \qquad \qquad + \big( 1 - s(t) \big) \big( 1 - \lambda(t) \big) \big( m_u^{\omega}(t) - m_d^{\omega}(t) \big) - h_u^{\omega}(t) m_u^{\omega}(t) - h_d^{\omega}(t) m_d^{\omega}(t) \Big) \Bigg] \\
&\quad \cdot \sum_{\{ \sigma_j^{\omega}(t) \}} \exp{\Bigg[ \sum_j \bigg( i \int_0^{\tau} \textrm{d}t \sum_{\omega} \omega \Big( \delta_{a_j,1} h_u^{\omega}(t) \sigma_j^{\omega}(t) + \delta_{a_j,-1} h_d^{\omega}(t) \sigma_j^{\omega}(t) \Big) + H_{\textrm{TF},j} \bigg) \Bigg]} \prod_j \big< \sigma_j^+(0) \big| \rho_j \big| \sigma_j^-(0) \big>.
\end{aligned}
\end{equation}
The sum over spin configurations now factors over $j$, and once again, there are only two distinct factors depending on whether $a_j = 1$ or $-1$ (recall that $\rho_j$ is the marked state $| a_j \rangle \langle a_j |$):
\begin{equation} \label{eq:Keldysh_generating_functional_single_spin}
\begin{aligned}
\mathcal{Z}_u &\equiv \sum_{\{ \sigma^{\omega}(t) \}} \exp{\Bigg[ i \int_0^{\tau} \textrm{d}t \sum_{\omega} \omega h_u^{\omega}(t) \sigma^{\omega}(t) + H_{\textrm{TF}} \Bigg]} \delta_{\sigma^+(0),1} \delta_{\sigma^-(0),1}, \\
\mathcal{Z}_d &\equiv \sum_{\{ \sigma^{\omega}(t) \}} \exp{\Bigg[ i \int_0^{\tau} \textrm{d}t \sum_{\omega} \omega h_d^{\omega}(t) \sigma^{\omega}(t) + H_{\textrm{TF}} \Bigg]} \delta_{\sigma^+(0),-1} \delta_{\sigma^-(0),-1}.
\end{aligned}
\end{equation}
Thus $\mathcal{Z}$ amounts to the path integral of $\exp{[iNS]}$, with action
\begin{equation} \label{eq:Keldysh_action}
\begin{aligned}
S &= \int_0^{\tau} \textrm{d}t \sum_{\omega} \omega \Big( s(t) \big( m_u^{\omega}(t) + m_d^{\omega}(t) \big)^p + s(t) \alpha \big( m_u^{\omega}(t) - m_d^{\omega}(t) \big)^p \\
&\qquad \qquad \qquad \qquad \qquad \qquad + \big( 1 - s(t) \big) \big( 1 - \lambda(t) \big) \big( m_u^{\omega}(t) - m_d^{\omega}(t) \big) - h_u^{\omega}(t) m_u^{\omega}(t) - h_d^{\omega}(t) m_d^{\omega}(t) \Big) \\
&\qquad - i (1-x) \log{\mathcal{Z}_u(h_u^{\pm})} - ix \log{\mathcal{Z}_d(h_d^{\pm})}.
\end{aligned}
\end{equation}
We evaluate the path integral by saddle-point approximation.
The integration variables are $m_r^{\omega}(t)$ and $h_r^{\omega}(t)$ for $r \in \{u, d\}$ and $\omega \in \{+, -\}$, so there are eight saddle-point equations for each $t$:
\begin{equation} \label{eq:Keldysh_saddle_point_equations_v1}
\begin{gathered}
s(t) p \big( m_u^{\omega}(t) + m_d^{\omega}(t) \big)^{p-1} + s(t) \alpha p \big( m_u^{\omega}(t) - m_d^{\omega}(t) \big)^{p-1} + \big( 1 - s(t) \big) \big( 1 - \lambda(t) \big) = h_u^{\omega}(t), \\
s(t) p \big( m_u^{\omega}(t) + m_d^{\omega}(t) \big)^{p-1} - s(t) \alpha p \big( m_u^{\omega}(t) - m_d^{\omega}(t) \big)^{p-1} - \big( 1 - s(t) \big) \big( 1 - \lambda(t) \big) = h_d^{\omega}(t), \\
\omega m_u^{\omega}(t) = -i (1-x) \frac{\partial \log{\mathcal{Z}_u(h_u^{\pm})}}{\partial h_u^{\omega}(t)}, \qquad \omega m_d^{\omega}(t) = -ix \frac{\partial \log{\mathcal{Z}_d(h_d^{\pm})}}{\partial h_d^{\omega}(t)}.
\end{gathered}
\end{equation}
It is consistent to take all variables to be independent of $\omega$ (this is analogous to the static ansatz for the thermodynamics).
Then $\mathcal{Z}_u$ (similarly $\mathcal{Z}_d$) becomes the Trotterized generating functional for a single spin that evolves in Hamiltonian $H_{u(d)}(t) = -h_{u(d)}(t) \hat{\sigma}^z - (1 - s(t)) \lambda(t) \hat{\sigma}^x$, starting from the up (down) state.
Furthermore,
\begin{equation} \label{eq:Keldysh_saddle_point_derivative_average}
\begin{aligned}
\frac{\partial \log{\mathcal{Z}_u}}{\partial h_u^{\omega}(t)} &= i \omega \sum_{\{ \sigma^{\omega}(t) \}} \sigma^{\omega}(t) \exp{\Bigg[ i \int_0^{\tau} \textrm{d}t \sum_{\omega} \omega h_u(t) \sigma^{\omega}(t) + H_{\textrm{TF}} \Bigg]} \delta_{\sigma^+(0),1} \delta_{\sigma^-(0),1} \\
&= i \omega \textrm{Tr} \hat{\sigma}^z \Big( \mathcal{T} e^{-i \int_0^t \textrm{d}t' H_u(t')} \Big) \big| \uparrow \big> \big< \uparrow \big| \Big( \mathcal{T} e^{-i \int_0^t \textrm{d}t' H_u(t')} \Big)^{\dag} = i \omega \big< \hat{\sigma}^z(t) \big>_u,
\end{aligned}
\end{equation}
and similarly $\partial \log{\mathcal{Z}_d}/\partial h_d^{\omega}(t) = i \omega \big< \hat{\sigma}^z(t) \big>_d$ (we used that $\mathcal{Z}_{u(d)} = 1$ for a generating functional).
Here $\langle \hat{\sigma}^z(t) \rangle_{u(d)}$ denotes the expectation value of $\hat{\sigma}^z$ in the state that begins pointing up (down) and evolves until time $t$ under the Hamiltonian $H_{u(d)}(t')$.
The saddle-point equations amount to
\begin{equation} \label{eq:Keldysh_saddle_point_equations_v2}
\begin{gathered}
s(t) p \big( m_u(t) + m_d(t) \big)^{p-1} + s(t) \alpha p \big( m_u(t) - m_d(t) \big)^{p-1} + \big( 1 - s(t) \big) \big( 1 - \lambda(t) \big) = h_u(t), \\
s(t) p \big( m_u(t) + m_d(t) \big)^{p-1} - s(t) \alpha p \big( m_u(t) - m_d(t) \big)^{p-1} - \big( 1 - s(t) \big) \big( 1 - \lambda(t) \big) = h_d(t), \\
m_u(t) = (1-x) \big< \hat{\sigma}^z(t) \big>_u, \qquad m_d(t) = x \big< \hat{\sigma}^z(t) \big>_d.
\end{gathered}
\end{equation}
\end{widetext}

Eqs.~\eqref{eq:Keldysh_saddle_point_equations_v2} are to be solved self-consistently, and as discussed above, the solutions for $m_u(t)$ and $m_d(t)$ are the expectation values $\langle \hat{S}_u^z(t) \rangle/N$ and $\langle \hat{S}_d^z(t) \rangle/N$ in the original problem.
Due to causality, the equations are straightforward to solve.
The upper two equations give $h_u(t)$ and $h_d(t)$ in terms of $m_u(t)$ and $m_d(t)$, and the lower equations state that the latter are (up to factors of $1-x$ and $x$) the corresponding expectation values of $\hat{\sigma}^z$ at time $t$.
Those expectation values only depend on the fields at previous times, so once $h_u$ and $h_d$ have been determined up to time $t$, the states are evolved for an additional small $\Delta t$ and expectation values are taken to determine $h_u$ and $h_d$ at time $t + \Delta t$.
In short, the algorithm for determining $m_u(t)$ and $m_d(t)$ is as follows:
\begin{widetext}
\begin{enumerate}
\item Begin with a spin (call it spin $u$) in the state $| \psi_u(0) \rangle = | \uparrow \; \rangle$, and a spin (call it spin $d$) in the state $| \psi_d(0) \rangle = | \downarrow \; \rangle$.
\item Do the following in order, setting $t = 0$:
\begin{itemize}
\item Calculate $m_u(t) = (1-x) \langle \psi_u(t) | \hat{\sigma}^z | \psi_u(t) \rangle$ and $m_d(t) = x \langle \psi_d(t) | \hat{\sigma}^z | \psi_d(t) \rangle$.
\item Set
\begin{equation} \label{eq:Keldysh_self_consistent_fields}
\begin{aligned}
h_u(t) &= s(t) p \big( m_u(t) + m_d(t) \big)^{p-1} + s(t) \alpha p \big( m_u(t) - m_d(t) \big)^{p-1} + \big( 1 - s(t) \big) \big( 1 - \lambda(t) \big), \\
h_d(t) &= s(t) p \big( m_u(t) + m_d(t) \big)^{p-1} - s(t) \alpha p \big( m_u(t) - m_d(t) \big)^{p-1} - \big( 1 - s(t) \big) \big( 1 - \lambda(t) \big).
\end{aligned}
\end{equation}
\item Evolve $| \psi_u(t) \rangle$ for time $\Delta t$ under the Hamiltonian
\begin{equation} \label{eq:Keldysh_self_consistent_Hamiltonian_up}
H_u(t) = -h_u(t) \hat{\sigma}^z - \big( 1 - s(t) \big) \lambda(t) \hat{\sigma}^x,
\end{equation}
and evolve $| \psi_d(t) \rangle$ for time $\Delta t$ under the Hamiltonian
\begin{equation} \label{eq:Keldysh_self_consistent_Hamiltonian_down}
H_d(t) = -h_d(t) \hat{\sigma}^z - \big( 1 - s(t) \big) \lambda(t) \hat{\sigma}^x,
\end{equation}
giving the states $| \psi_u(t + \Delta t) \rangle$ and $| \psi_d(t + \Delta t) \rangle$.
\end{itemize}
\item Now that $| \psi_u(\Delta t) \rangle$ and $| \psi_d(\Delta t) \rangle$ have been determined, repeat step 2 for $t = \Delta t$ to obtain $| \psi_u(2 \Delta t) \rangle$ and $| \psi_d(2 \Delta t) \rangle$, then repeat again to obtain $| \psi_u(3 \Delta t) \rangle$ and $| \psi_d(3 \Delta t) \rangle$, and so on.
\end{enumerate}
\end{widetext}

\section{SRA dynamical path integral} \label{app:SRA_path_integral}

Recall that SRA consists of a Metropolis Monte Carlo simulation using the Hamiltonian
\begin{equation} \label{eq:simulated_reverse_annealing_Hamiltonian_repeat}
H(s, \lambda) = s H_0 - (1 - s)(1 - \lambda) \sum_{j=1}^N a_j \sigma_j,
\end{equation}
and temperature
\begin{equation} \label{eq:simulated_reverse_annealing_temperature_repeat}
T(s, \lambda) = (1 - s) \lambda,
\end{equation}
where $s(t)$ and $\lambda(t)$ vary during the course of the simulation (again denote the total runtime by $\tau$).
One certainly could use a more elaborate Monte Carlo algorithm in place of single-spin Metropolis updates, but this simple protocol is more amenable to analytical analysis.

While it is far less common than the Keldysh path integral, one can construct a path-integral representation of a generating functional for stochastic Metropolis Monte Carlo dynamics.
To begin, consider a generic classical Ising model, not necessarily Eq.~\eqref{eq:simulated_reverse_annealing_Hamiltonian_repeat}.
Given a spin configuration $\{ \sigma_j \}$, define the local field $b_j$ by calculating the change in energy $\Delta E$ upon flipping spin $j$ and setting $\Delta E = 2 b_j \sigma_j$.
Note that $b_j$ depends on the overall spin configuration, so it is technically a function $b_j(\{ \sigma_k \})$, but we do not indicate this to keep the notation more manageable.
We assume that each spin evolves independently at a rate given by the Metropolis update rule, which is itself determined by the spin's local field.
More precisely, we choose a small (ideally infinitesimal) timestep $\Delta t$, and take the probability of the spin configuration changing from $\{ \sigma_j \}$ to $\{ \sigma'_j \}$ during the timestep to be $\prod_j \pi(\sigma'_j | \sigma_j; b_j)$, with
\begin{equation} \label{eq:Metropolis_update_rule}
\pi \big( \sigma'_j | \sigma_j; b_j \big) = \begin{cases} 1 - \gamma \Delta t \min \big\{ e^{-2 \beta b_j \sigma_j}, 1 \big\}, \; &\sigma'_j = \sigma_j \\ \gamma \Delta t \min \big\{ e^{-2 \beta b_j \sigma_j}, 1 \big\}, \; &\sigma'_j = -\sigma_j \end{cases}.
\end{equation}
This amounts to, independently for each spin, first deciding whether to attempt a flip with probability $\gamma \Delta t$ (where $\gamma$ is an overall constant rate) and then actually flipping the spin with probability $\min \{ e^{-2 \beta b_j \sigma_j}, 1 \}$ if so.
Technically this is not what the Monte Carlo simulation does --- there one only attempts to flip a single spin at a time --- but the difference is negligible for sufficiently small $\Delta t$, where the probability of attempting multiple flips simultaneously is asymptotically smaller.

The stochastic generating functional is
\begin{equation} \label{eq:stochastic_generating_functional_definition}
\mathcal{Z} \equiv \sum_{\{ \sigma_j(t) \}} \prod_{t,j} \pi \big( \sigma_j(t + \Delta t) \big| \sigma_j(t); b_j(t) \big) \prod_j \rho_{0j} \big( \sigma_j(0) \big),
\end{equation}
where $\rho_{0j}(\sigma_j(0))$ is the probability distribution of the initial state $\sigma_j(0)$.
The summand is simply the probability of having a specific sequence of configurations $\{ \sigma_j(0) \} \rightarrow \{ \sigma_j(\Delta t) \} \rightarrow \{ \sigma_j(2 \Delta t) \} \rightarrow \cdots$, and so summing over all sequences of configurations means that $\mathcal{Z} = 1$, exactly as for the Keldysh generating functional.
Yet once again, we consider $\mathcal{Z}$ because its path-integral representation will involve an integral over variables $m_u(t)$ and $m_d(t)$ whose saddle-point values are the expectation values of $S_u(t) \equiv \sum_j \delta_{a_j,1} \sigma_j(t)$ and $S_d(t) \equiv \sum_j \delta_{a_j,-1} \sigma_j(t)$.
We aim to derive an equation determining that saddle point.

Now specialize to Eq.~\eqref{eq:simulated_reverse_annealing_Hamiltonian_repeat}, first by computing the local fields.
The spin configuration $\{ \sigma_k \}$ has energy
\begin{widetext}
\begin{equation} \label{eq:SRA_energy_function}
E \big( \{ \sigma_k \}, t \big) = -N s(t) \bigg( \frac{1}{N} \sum_k \sigma_k \bigg)^p - N s(t) \alpha \bigg( \frac{1}{N} \sum_k a_k \sigma_k \bigg)^p - \big( 1-  s(t) \big) \big( 1 - \lambda(t) \big) \sum_k a_k \sigma_k,
\end{equation}
and so if we take $\sigma_j \rightarrow -\sigma_j = \sigma_j - 2 \sigma_j$, the energy changes by
\begin{equation} \label{eq:SRA_energy_change_function}
\begin{aligned}
\Delta E &= -N s(t) \bigg( \frac{1}{N} \sum_k \sigma_k - \frac{2 \sigma_j}{N} \bigg)^p + N s(t) \bigg( \frac{1}{N} \sum_k \sigma_k \bigg)^p - N s(t) \alpha \bigg( \frac{1}{N} \sum_k a_k \sigma_k - \frac{2 a_j \sigma_j}{N} \bigg)^p + N s(t) \alpha \bigg( \frac{1}{N} \sum_k a_k \sigma_k \bigg)^p \\
&\qquad + 2 \big( 1 - s(t) \big) \big( 1 - \lambda(t) \big) a_j \sigma_j \\
&\sim 2 s(t) p \bigg( \frac{1}{N} \sum_k \sigma_k \bigg)^{p-1} \sigma_j + 2 a_j s(t) \alpha p \bigg( \frac{1}{N} \sum_k a_k \sigma_k \bigg)^{p-1} \sigma_j + 2 a_j \big( 1 - s(t) \big) \big( 1 - \lambda(t) \big) \sigma_j,
\end{aligned}
\end{equation}
where we Taylor-expanded to obtain the bottom line (all higher-order terms vanish at large $N$).
Equating this to $2 b_j(t) \sigma_j$, the local field at time $t$ is
\begin{equation} \label{eq:SRA_local_field}
b_j(t) = s(t) p \bigg( \frac{1}{N} \sum_k \sigma_k \bigg)^{p-1} + a_j s(t) \alpha p \bigg( \frac{1}{N} \sum_k a_k \sigma_k \bigg)^{p-1} + a_j \big( 1 - s(t) \big) \big( 1 - \lambda(t) \big).
\end{equation}
Note that $b_j(t)$ depends on the spin configuration only through the quantities $\sum_k \sigma_k$ and $\sum_k a_k \sigma_k$, equivalently $\sum_k \delta_{a_k,1} \sigma_k$ and $\sum_k \delta_{a_k,-1} \sigma_k$.
Thus in the generating functional, introduce $\delta$-functions fixing $\sum_k \delta_{a_k,1} \sigma_k(t)$ to equal $Nm_u(t)$ and $\sum_k \delta_{a_k,-1} \sigma_k(t)$ to equal $Nm_d(t)$.
Then $b_j(t)$ takes only one of two values: either (when $a_j = 1$)
\begin{equation} \label{eq:stochastic_mean_local_field_up}
\overline{b}_u(t) \equiv s(t) p \big( m_u(t) + m_d(t) \big)^{p-1} + s(t) \alpha p \big( m_u(t) - m_d(t) \big)^{p-1} + \big( 1 - s(t) \big) \big( 1 - \lambda(t) \big),
\end{equation}
or (when $a_j = -1$)
\begin{equation} \label{eq:stochastic_mean_local_field_down}
\overline{b}_d(t) \equiv s(t) p \big( m_u(t) + m_d(t) \big)^{p-1} - s(t) \alpha p \big( m_u(t) - m_d(t) \big)^{p-1} - \big( 1 - s(t) \big) \big( 1 - \lambda(t) \big).
\end{equation}
We write $b_j(t) = \delta_{a_j,1} \overline{b}_u(t) + \delta_{a_j,-1} \overline{b}_d(t)$, and the generating functional becomes
\begin{equation} \label{eq:stochastic_generating_functional_v2}
\begin{aligned}
\mathcal{Z} &= \int \prod_t \textrm{d}m_u(t) \textrm{d}m_d(t) \sum_{\{ \sigma_j(t) \}} \prod_{t,j} \pi \big( \sigma_j(t + \Delta t) \big| \sigma_j(t); \delta_{a_j,1} \overline{b}_u(t) + \delta_{a_j,-1} \overline{b}_d(t) \big) \\
&\qquad \qquad \qquad \qquad \qquad \qquad \cdot \prod_t \delta \bigg( Nm_u(t) - \sum_j \delta_{a_j,1} \sigma_j(t) \bigg) \delta \bigg( Nm_d(t) - \sum_j \delta_{a_j,-1} \sigma_j(t) \bigg) \prod_j \rho_{0j} \big( \sigma_j(0) \big).
\end{aligned}
\end{equation}
Again express the $\delta$-functions as integrals of complex exponentials:
\begin{equation} \label{eq:stochastic_generating_functional_v3}
\begin{aligned}
\mathcal{Z} &= \int \prod_t \textrm{d}m_u(t) \textrm{d}m_d(t) \textrm{d}h_u(t) \textrm{d}h_d(t) \exp{\Bigg[ i N \int_0^{\tau} \textrm{d}t \Big( h_u(t) m_u(t) + h_d(t) m_d(t) \Big) \Bigg]} \\
&\qquad \cdot \sum_{\{ \sigma_j(t) \}} \prod_{t,j} \pi \big( \sigma_j(t + \Delta t) \big| \sigma_j(t); \delta_{a_j,1} \overline{b}_u(t) + \delta_{a_j,-1} \overline{b}_d(t) \big) \exp{\Bigg[ -i \sum_j \int_0^{\tau} \textrm{d}t \Big( \delta_{a_j,1} h_u(t) \sigma_j(t) + \delta_{a_j,-1} h_d(t) \sigma_j(t) \Big) \Bigg]} \\
&\qquad \qquad \qquad \qquad \cdot \prod_j \rho_{0j} \big( \sigma_j(0) \big),
\end{aligned}
\end{equation}
and the sum over spin configurations now factors over $j$ (whereas it did not originally because the local fields depend on all $\{ \sigma_k \}$ --- here they have been replaced by separate integration variables).
The two possible factors are
\begin{equation} \label{eq:stochastic_generating_functional_single_spin}
\begin{aligned}
\mathcal{Z}_u &\equiv \sum_{\{ \sigma(t) \}} \prod_t \pi \big( \sigma(t + \Delta t) \big| \sigma(t); \overline{b}_u(t) \big) \exp{\Bigg[ -i \int_0^{\tau} \textrm{d}t \, h_u(t) \sigma(t) \Bigg]} \delta_{\sigma(0),1}, \\
\mathcal{Z}_d &\equiv \sum_{\{ \sigma(t) \}} \prod_t \pi \big( \sigma(t + \Delta t) \big| \sigma(t); \overline{b}_d(t) \big) \exp{\Bigg[ -i \int_0^{\tau} \textrm{d}t \, h_d(t) \sigma(t) \Bigg]} \delta_{\sigma(0),-1},
\end{aligned}
\end{equation}
and the full generating functional amounts to the path integral of $\exp{[iNS]}$ with action
\begin{equation} \label{eq:stochastic_action}
S[m_u, m_d, h_u, h_d] = \int_0^{\tau} \textrm{d}t \Big( h_u(t) m_u(t) + h_d(t) m_d(t) \Big) - i (1-x) \log{\mathcal{Z}_u[m_u, m_d, h_u]} - ix \log{\mathcal{Z}_d[m_u, m_d, h_d]}.
\end{equation}
Here we have explicitly indicated which of the integration variables ($m_u$, $m_d$, $h_u$, $h_d$) each of the terms depends on.
We can now carry out the integrals by saddle-point approximation, giving four equations at each time:
\begin{equation} \label{eq:stochastic_saddle_point_equations_v1}
\begin{aligned}
h_u(t) &= i (1-x) \frac{\partial \log{\mathcal{Z}_u[m_u, m_d, h_u]}}{\partial m_u(t)} + ix \frac{\partial \log{\mathcal{Z}_d[m_u, m_d, h_d]}}{\partial m_u(t)}, \quad &m_u(t) &= i (1-x) \frac{\partial \log{\mathcal{Z}_u[m_u, m_d, h_u]}}{\partial h_u(t)}, \\
h_d(t) &= i (1-x) \frac{\partial \log{\mathcal{Z}_u[m_u, m_d, h_u]}}{\partial m_d(t)} + ix \frac{\partial \log{\mathcal{Z}_d[m_u, m_d, h_d]}}{\partial m_d(t)}, \quad &m_d(t) &= ix \frac{\partial \log{\mathcal{Z}_d[m_u, m_d, h_d]}}{\partial h_d(t)}.
\end{aligned}
\end{equation}
A consistent solution to these equations has $h_u(t) = h_d(t) = 0$.
To see this, note that regardless of how $m_u(t)$ and $m_d(t)$ depend on time, $\mathcal{Z}_u[m_u, m_d, 0]$ is the generating functional for a single spin in time-dependent field $\overline{b}_u(t)$ and hence equal to 1:
\begin{equation} \label{eq:stochastic_single_spin_simplification}
\mathcal{Z}_u[m_u, m_d, 0] \equiv \sum_{\{ \sigma(t) \}} \prod_t \pi \big( \sigma(t + \Delta t) \big| \sigma(t); \overline{b}_u(t) \big) \delta_{\sigma(0),1} = 1.
\end{equation}
In other words, $\mathcal{Z}_u$ is independent of $m_u(t)$ and $m_d(t)$ at $h_u = 0$, and so $\partial \mathcal{Z}_u/\partial m_u(t) = \partial \mathcal{Z}_u/\partial m_d(t) = 0$.
The same holds for the derivatives of $\mathcal{Z}_d[m_u, m_d, 0]$.
Thus the left-hand equations in Eq.~\eqref{eq:stochastic_saddle_point_equations_v1} become $0 = 0$ upon setting $h_u$ and $h_d$ to zero.
As for the right-hand equations, following the same steps,
\begin{equation} \label{eq:stochastic_single_spin_derivative_average}
i \frac{\partial \log{\mathcal{Z}_u[m_u, m_d, h_u]}}{\partial h_u(t)} \Bigg|_{h_u = 0} = \sum_{\{ \sigma(t') \}} \sigma(t) \prod_{t'} \pi \big( \sigma(t' + \Delta t) \big| \sigma(t'); \overline{b}_u(t') \big) \delta_{\sigma(0),1} = \big< \sigma(t) \big>_u,
\end{equation}
\end{widetext}
where $\langle \sigma(t) \rangle_u$ denotes the expectation value of $\sigma(t)$ for the single spin in local field $\overline{b}_u(t)$, starting from the up state.
Similarly, $i \partial \log{\mathcal{Z}_d}/\partial h_d(t)|_{h_d = 0} = \langle \sigma(t) \rangle_d$, where the right-hand side is the expectation value in local field $\overline{b}_d(t)$ starting from the down state.
The remaining saddle-point equations thus amount to
\begin{equation} \label{eq:stochastic_saddle_point_equations_v2}
m_u(t) = (1-x) \big< \sigma(t) \big>_u, \qquad m_d(t) = x \big< \sigma(t) \big>_d.
\end{equation}
To solve them, we need only run two single-spin Monte Carlo simulations: one for a spin that starts pointing up and experiences field $\overline{b}_u(t)$, and another for a spin that starts pointing down and experiences field $\overline{b}_d(t)$.
Upon computing $\langle \sigma(t) \rangle_u$ and $\langle \sigma(t) \rangle_d$, we set $m_u(t) = (1-x) \langle \sigma(t) \rangle_u$ and $m_d(t) = x \langle \sigma(t) \rangle_d$.
This determines $\overline{b}_u(t)$ and $\overline{b}_d(t)$, which allows us to advance the simulation to time $t + \Delta t$ --- in this way, we determine $m_u(t)$ and $m_d(t)$ at all times.

However, we can actually be more efficient.
Rather than explicitly running a simulation and computing $\langle \sigma(t) \rangle_{u(d)}$ by taking a number of samples, suppose that we have determined the exact probability distribution $\rho(\sigma; t)$.
Since the simulation involves only a single spin, $\rho(\sigma; t)$ consists of the two values $\rho(\uparrow; t)$ and $\rho(\downarrow; t)$ (which are themselves related by normalization).
Eq.~\eqref{eq:Metropolis_update_rule} is the transition matrix giving $\rho(\sigma'; t + \Delta t)$ in terms of $\rho(\sigma; t)$, i.e.,
\begin{equation} \label{eq:stochastic_single_spin_probability_update}
\rho(\sigma'; t + \Delta t) = \sum_{\sigma} \pi \big( \sigma' \big| \sigma; \overline{b}(t) \big) \rho(\sigma; t),
\end{equation}
and since we know the initial distribution ($\rho(\sigma; 0) = \delta_{\sigma,1}$ for spin $u$ and $\rho(\sigma; 0) = \delta_{\sigma,-1}$ for spin $d$), we can determine $\rho(\sigma; t)$ for all $t$.
Then $\langle \sigma(t) \rangle = \rho(\uparrow; t) - \rho(\downarrow; t)$ by definition.
Keep in mind that we do this separately for spins $u$ and $d$, with different fields $\overline{b}_u(t)$ and $\overline{b}_d(t)$.

To summarize, the algorithm for determining $m_u(t)$ and $m_d(t)$ is as follows:
\begin{widetext}
\begin{enumerate}
\item Begin with a spin (call it spin $u$) having distribution $\rho_u(\sigma; 0) = \delta_{\sigma,1}$, and a spin (call it spin $d$) having distribution $\rho_d(\sigma; 0) = \delta_{\sigma,-1}$.
\item Do the following in order, setting $t = 0$:
\begin{itemize}
\item Calculate $m_u(t) = (1 - x) [\rho_u(\uparrow; t) - \rho_u(\downarrow; t)]$ and $m_d(t) = x [\rho_d(\uparrow; t) - \rho_d(\downarrow; t)]$.
\item Set
\begin{equation} \label{eq:stochastic_mean_local_fields}
\begin{aligned}
\overline{b}_u(t) &= s(t) p \big( m_u(t) + m_d(t) \big)^{p-1} + s(t) \alpha p \big( m_u(t) - m_d(t) \big)^{p-1} + \big( 1 - s(t) \big) \big( 1 - \lambda(t) \big), \\
\overline{b}_d(t) &= s(t) p \big( m_u(t) + m_d(t) \big)^{p-1} - s(t) \alpha p \big( m_u(t) - m_d(t) \big)^{p-1} - \big( 1 - s(t) \big) \big( 1 - \lambda(t) \big).
\end{aligned}
\end{equation}
\item Compute
\begin{equation} \label{eq:stochastic_distribution_updates}
\rho_u(\sigma'; t + \Delta t) = \sum_{\sigma} \pi \big( \sigma' \big| \sigma; \overline{b}_u(t) \big) \rho_u(\sigma; t), \qquad \rho_d(\sigma'; t + \Delta t) = \sum_{\sigma} \pi \big( \sigma' \big| \sigma; \overline{b}_d(t) \big) \rho_d(\sigma; t),
\end{equation}
where $\pi(\sigma' | \sigma; \overline{b}(t))$ is given by
\begin{equation} \label{eq:Metropolis_update_rule_repeat}
\pi \big( \sigma' | \sigma; \overline{b}(t) \big) = \begin{cases} 1 - \gamma \Delta t \min \big\{ e^{-2 \beta(t) \overline{b}(t) \sigma}, 1 \big\}, \; &\sigma' = \sigma \\ \gamma \Delta t \min \big\{ e^{-2 \beta(t) \overline{b}(t) \sigma}, 1 \big\}, \; &\sigma' = -\sigma \end{cases}.
\end{equation}
\end{itemize}
\item Now that $\rho_u(\sigma; \Delta t)$ and $\rho_d(\sigma; \Delta t)$ have been determined, repeat step 2 for $t = \Delta t$ to obtain $\rho_u(\sigma; 2 \Delta t)$ and $\rho_d(\sigma; 2 \Delta t)$, then repeat again to obtain $\rho_u(\sigma; 3 \Delta t)$ and $\rho_d(\sigma; 3 \Delta t)$, and so on.
\end{enumerate}
\end{widetext}

\bibliography{biblio}

\begin{thebibliography}{60}
\bibitem [{sta()}]{static_ansatz_note}%
  \BibitemOpen
  \href@noop {} {}\bibinfo {note} {The static ansatz is equivalently the assumption that (imaginary-)time-translation symmetry is not spontaneously broken, which is a very reasonable assumption to make.}\BibitemShut {Stop}%
\bibitem [{the()}]{thermal_expectation_note}%
  \BibitemOpen
  \href@noop {} {}\bibinfo {note} {One can see this by following the same derivation of the path integral, but for the thermal expectation value of $\sum_j \delta_{a_j,1} \hat{\sigma}_j^z/N$ (similarly for $\sum_j \delta_{a_j,-1} \hat{\sigma}_j^z/N$) rather than the partition function. The result is the path integral of $m_u \exp{[-N \beta \Phi]}/Z$, which according to the saddle-point approximation is simply $m_u$ evaluated at the saddle point (the factor of $\exp{[-N \beta \Phi]}$ cancels with $Z$ in the denominator).}\BibitemShut {Stop}%
\bibitem [{spi()}]{spinodal_note}%
  \BibitemOpen
  \href@noop {} {}\bibinfo {note} {This statement is more subtle than it may appear: even though both local minima are stable on either side of a discontinuous phase transition, there can be separate ``spinodal'' lines at which the secondary minimum becomes unstable. If the spinodal is crossed, there is no longer a (free) energy barrier and the system may be able to reach the desired ground state, especially via SRA since it directly moves downhill in free energy. The figure uses parameter values for which there is no such spinodal.}\BibitemShut {Stop}%
\end{thebibliography}

\end{document}